\documentclass[english,journal,12pt,onecolumn,draftclsnofoot]{IEEEtran}
\usepackage[T1]{fontenc}
\usepackage[latin9]{inputenc}
\usepackage{color}
\usepackage{float}
\usepackage{amsmath}
\usepackage{amssymb}
\usepackage{graphicx}

\makeatletter

\floatstyle{ruled}
\newfloat{algorithm}{tbp}{loa}
\providecommand{\algorithmname}{Algorithm}
\floatname{algorithm}{\protect\algorithmname}

\usepackage{flushend}

\@ifundefined{showcaptionsetup}{}{%
 \PassOptionsToPackage{caption=false}{subfig}}
\usepackage{subfig}
\makeatother

\usepackage{babel}
\begin{document}
\title{Optimum Low-Complexity Decoder for Spatial Modulation}
\author{Ibrahim Al-Nahhal, \textit{Student Member, IEEE}, Ertugrul Basar,
\textit{Senior Member, IEEE}, Octavia A. Dobre, \textit{Senior Member,
IEEE}, and Salama Ikki, \textit{Senior Member, IEEE}}
\maketitle
\begin{abstract}
In this paper, a novel low-complexity detection algorithm for spatial
modulation (SM), referred to as the minimum-distance of maximum-length
(m-M) algorithm, is proposed and analyzed. The proposed m-M algorithm
is a smart searching method that is applied for the SM tree-search
decoders. The behavior of the m-M algorithm is studied for three different
scenarios: i) perfect channel state information at the receiver side
(CSIR), ii) imperfect CSIR of a fixed channel estimation error variance,
and iii) imperfect CSIR of a variable channel estimation error variance.
Moreover, the complexity of the m-M algorithm is considered as a random
variable, which is carefully analyzed for all scenarios, using probabilistic
tools. Based on a combination of the sphere decoder (SD) and ordering
concepts, the m-M algorithm guarantees to find the maximum-likelihood
(ML) solution with a significant reduction in the decoding complexity
compared to SM-ML and existing SM-SD algorithms; it can reduce the
complexity up to $94\%$ and $85\%$ in the perfect CSIR and the worst
scenario of imperfect CSIR, respectively, compared to the SM-ML decoder.
Monte Carlo simulation results are provided to support our findings
as well as the derived analytical complexity reduction expressions.
\end{abstract}


\noindent {\footnotesize{}This work is supported by the Natural Sciences
and Engineering Research Council of Canada (NSERC), through its Discovery
program. The work of E. Basar is supported in part by the Turkish
Academy of Sciences (TUBA), GEBIP Programme. This work was presented
in part at IEEE VTC Spring 2018, Portugal {[}\ref{I.-Al-Nahhal,-O. VTC}{]}.}{\footnotesize\par}

{\footnotesize{}O. A. Dobre and I. Al-Nahhal are with the Faculty
of Engineering and Applied Science, Memorial University, 300 Prince
Phillip Dr., St. John\textquoteright s, NL, A1B 3X5, Canada (e-mail:
\{odobre, ioalnahhal\}@mun.ca).}{\footnotesize\par}

{\footnotesize{}E. Basar is with the Communications Research and Innovation
Laboratory (CoreLab), Department of Electrical and Electronics Engineering,
Ko\c{c}\c{ } University, Sariyer 34450, Istanbul, Turkey (e-mail:
ebasar@ku.edu.tr).}{\footnotesize\par}

{\footnotesize{}S. Ikki is with the Department of Electrical Engineering,
Lakehead University, Thunder Bay, ON P7B 5E1, Canada (e-mail: sikki@lakeheadu.ca).}{\footnotesize\par}

\section{Introduction}

\def\figurename{Fig.}
\def\tablename{TABLE}

\IEEEPARstart{M}{ultiple}-input multiple-output (MIMO) systems, which
is an integral part of modern wireless communication standards, activate
all transmit antennas to increase the spectral efficiency and/or improve
the bit-error-ratio (BER) performance {[}\ref{E.-Telatar,-=00201CCapacity}{]}.
On the other hand, activating all transmit antennas at the same time
not only creates a strong inter-channel interference (ICI) but also
requires multiple radio frequency chains. A promising technique called
spatial modulation (SM) has been studied in recent years {[}\ref{C.-X.-Wang-et}{]}-{[}\ref{E.-Basar,-=00201CIndex}{]}
to overcome these problems in next-generation systems. In SM {[}\ref{R.-Mesleh,-H.}{]}-{[}\ref{M.-Di-Renzo, spatial modulation for multiple}{]},
only one transmit antenna is activated during the transmission burst,
where the active transmit antenna is chosen out of all transmit antennas
according to a part of the input bit-stream. The active antenna transmits
a phase shift keying (PSK) or quadrature amplitude modulation (QAM)
symbol, through a wireless medium, based on the rest of the input
bit-stream. At the receiver side, all receive antennas receive the
delivered signal and forward it to the digital signal processor (DSP)
unit for decoding. The maximum-likelihood (ML) detector is utilized
to decode the received signal by attempting all possible combinations
of the QAM/PSK symbols and the transmit antennas, where this process
depends on the number of transmit antennas, receive antennas, and
modulation order. Consequently, the ML algorithm is classified to
be costly from the decoding complexity point of view, particularly
for increasing number of transmit/receive antennas and constellation
points. 

Low-latency communications and energy-efficient transmission techniques
are among the next generation (5G) requirements {[}\ref{J.-G.-Andrews,}{]};
one solution to achieve this is the design of low-complexity decoding
algorithms for the SM system. Recently, low-complexity decoding algorithms
have been proposed for the SM system in {[}\ref{A.-Younis,-R.GLOBOCOM}{]}-{[}\ref{X.-Zhang,-Y.}{]},
and surveyed in {[}\ref{P.-Yang,-M.}{]}. In {[}\ref{A.-Younis,-R.GLOBOCOM}{]}-{[}\ref{A.-Younis,-S. IEEE Trans}{]},
the sphere decoding (SD) concept of {[}\ref{E.-Viterbo-and 1999}{]},
{[}\ref{B.-Hassibi-and SD2005}{]} is exploited to provide a low-complexity
detection at the BER level of the brute-force ML detector. The authors
of {[}\ref{A.-Younis,-R.GLOBOCOM}{]}-{[}\ref{A.-Younis,-S. IEEE Trans}{]}
have provided a threshold (pruned radius for the SD) that depends
on the number of receive antennas, noise variance, and a predetermined
constant, which changes for each different MIMO system. The noise
variance estimation process is an exhaustive step required for every
change in the channel environment; it can be achieved either blindly
or using data-aided (DA) techniques like preamble/pilots {[}\ref{A.-Das-and SNR est}{]}-{[}\ref{F.-Bellili,-R. SNR est}{]}
transmission. In {[}\ref{Q.-Tang,-Y.}{]}, the authors have proposed
an algorithm that provides a trade-off between the BER performance
and decoding complexity for the SM decoders. This algorithm requires
an exhaustive pre-processing step to calculate the pseudo-inverse
of the channel matrix columns. This step is mitigated in {[}\ref{L.-Xiao,-P.}{]}
by considering a sparse channel of a large-scale MIMO system. However,
the problem of noise variance dependency still exists in {[}\ref{L.-Xiao,-P.}{]}.
Furthermore, the ML BER performance has not been achieved in {[}\ref{Q.-Tang,-Y.}{]}
and {[}\ref{L.-Xiao,-P.}{]}. The authors of {[}\ref{I.-Al-Nahhal,-O. QSM}{]}
have provided a low-complexity algorithm with the ML BER performance
for the quadrature SM (QSM) decoders by treating the QSM symbol as
two independent SM symbols. The reduction in the decoding complexity
comes from the ordering concept, with no dependency on the noise variance.
However, further reduction in the decoding complexity can be attained.
The authors in {[}\ref{X.-Zhang,-Y.}{]} have proposed an algorithm
with near-ML performance, which reduces the computational complexity
of the SM decoders based on modified beam search and ordering concepts,
by splitting the tree-search into sub-trees. It should be noted that
the algorithms in {[}\ref{A.-Younis,-R.GLOBOCOM}{]}-{[}\ref{X.-Zhang,-Y.}{]}
consider  perfect knowledge of the channel state information at the
receiver side (CSIR), and no study is presented in the case of imperfect
CSIR.

In this paper, we propose a low-complexity algorithm for the SM decoders,
referred to as the \textit{minimum-distance of maximum-length} (m-M)
algorithm. Based on the tree-search concept, the m-M algorithm performs
only one expansion to the minimum Euclidean distance (ED) across all
tree-search branches until the minimum ED occurs at the end of a fully
expanded branch. The proposed m-M algorithm provides a significant
reduction in the decoding complexity with the ML BER performance,
and requires no knowledge of the noise variance. We provide a complete
study of our proposed algorithm in the case of perfect and imperfect
CSIR. In case of imperfect CSIR, we consider two scenarios for the
fixed and variable variance of the error in the channel estimation,
respectively. In addition, we derive tight probabilistic expressions
for the expected decoding complexity of the m-M algorithm for all
scenarios.

The rest of the paper\footnote{Notations: Boldface uppercase and lowercase letters represent matrices
and vectors, respectively. $\mathcal{C}\mathcal{N}$ stands for a
complex-valued normally distributed random variable. $\left\Vert \centerdot\right\Vert $
denotes the Euclidean norm. $\left|\centerdot\right|$ returns the
absolute value of an element. $\centerdot^{\Re}$ and $\centerdot^{\Im}$
denote the real and imaginary components, respectively. $\mathbb{E}\left\{ \centerdot\right\} $
denotes the expectation operation. $\mathbb{P}\text{r}(\centerdot)$
is the probability of an event. $f_{\centerdot}(\centerdot)$ denotes
the probability density function (pdf) of a random variable. $\text{sum}\left\{ \centerdot\right\} $
returns the summation of all elements values of a vector. $k!$ stands
for the factorial operation of an integer $k$.} is organized as follows: In Section \ref{sec:System-Model}, the
system model of the SM transmitter and receiver is summarized. In
Section \ref{sec:Minimum-Distance-Maximum-Length-}, the proposed
m-M algorithm is introduced. In Section \ref{sec:Complexity-Analysis},
tight analytical expressions of the m-M algorithm decoding complexity
are derived for perfect and imperfect CSIR. In Section \ref{sec:Optimality-of-BER-Performance},
the optimality of the m-M algorithm is discussed. The numerical results
and conclusion are provided in Sections \ref{sec:Numerical-results}
and \ref{sec:Conclusion}, respectively. 

\section{\label{sec:System-Model}System Model}

\subsection{SM Modulator}

Consider the implementation of an SM scheme for $N_{r}\times N_{t}$
MIMO system, where $N_{t}$ and $N_{r}$ denote the number of transmit
and receive antennas, respectively. The incoming bit-stream is divided
into two groups: the first group of $\text{log}_{2}\left(N_{t}\right)$
bits selects the transmit antenna that will be activated, while the
second group of $\text{log}_{2}\left(M\right)$ bits selects the QAM/PSK
symbol that will be delivered from that antenna, where $M$ denotes
the order of the QAM/PSK constellation. Therefore, the number of bits
delivered in every time instance by the SM system is

\begin{equation}
\eta=\text{log}_{2}\left(N_{t}\right)+\text{log}_{2}\left(M\right),\label{eq: Spectral Efficiency}
\end{equation}

\noindent where $\eta$ denotes the spectral efficiency in bits per
channel use (bpcu). The active antenna transmits $s_{t}\in\left\{ s_{1},\ldots,s_{M}\right\} $
through a Rayleigh fading path between the transmit antenna and all
$N_{r}$ receive antennas, where $s_{t}$ is the transmitted QAM/PSK
symbol. This path represents the transmit channel, $\mathbf{h}_{t}\sim\mathcal{C}\mathcal{N}\left(0,1\right)$,
which is drawn from the full channel matrix, $\mathbf{H}\in\mathbb{C}^{N_{r}\times N_{t}}$.

Assume that the data symbol $s_{t}$ is transmitted over $\mathbf{h}_{t}$
to form the transmitted SM symbol combination, $\mathbf{x}_{t}\in\left\{ \mathbf{x}_{1},\ldots,\mathbf{x}_{MN_{t}}\right\} $,
where $\mathbf{x}_{t}=\mathbf{h}_{t}s_{t}$. It should be noted that
the transmitted combination is drawn from $MN_{t}$ different possible
combinations, which result from combining $M$ QAM/PSK symbols with
$N_{t}$ transmit antennas. Due to the additive white Gaussian noise
(AWGN), the SM symbol is received as 

\begin{equation}
\mathbf{y}=\mathbf{x}_{t}+\mathbf{w},\label{eq: y}
\end{equation}

\noindent where $\mathbf{y}\in\mathbb{C}^{N_{r}\times1}$ denotes
the noisy received vector and $\mathbf{w}\in\mathbb{C}^{N_{r}\times1}$
is the AWGN vector with entries having zero-mean and variance $\sigma_{n}^{2}$
(i.e., $\mathbf{w}\sim\mathcal{C}\mathcal{N}\left(0,\sigma_{n}^{2}\right)$).
Note that QAM is considered in this paper.

\subsection{SM-ML Demodulation}

At the receiver side, the DSP unit utilizes the ML detection algorithm
to estimate the transmitted combination. The ML algorithm attempts
all possible combinations to find the one that provides the minimum
ED with the received signal vector {[}\ref{J.-Jeganathan,-A.}{]},
which corresponds to the index of

\begin{equation}
\hat{j}_{\text{ML}}=\underset{j=1,\cdots,MN_{t}}{\text{arg}\,\text{min}}\left\Vert \mathbf{y}-\mathbf{x}_{j}\right\Vert ^{2}=\underset{j=1,\cdots,MN_{t}}{\text{arg}\,\text{min}}\sum_{n=1}^{N_{r}}\left|y_{n}-x_{n,j}\right|^{2},\label{eq: x_est_ML_index}
\end{equation}

\noindent where $\hat{j}_{\text{ML}}$ is the index of the estimated
combination using the ML detection algorithm, $y_{n}$ is the $n$-th
element of $\mathbf{y}$, and $x_{n,j}$ is the $n$-th element of
the $j$-th combination. 

\begin{figure}
\begin{centering}
\includegraphics[scale=0.47]{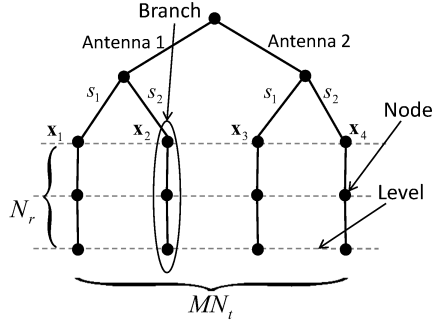}
\par\end{centering}
\caption{{\small{}\label{fig:SM-tree-search.}SM tree-search decoder for $M=2$,
$N_{t}=2$, and $N_{r}=3$ with four branches.}}

\end{figure}

It should be noted that estimating the transmitted combination can
be achieved using a graphical approach, named \textit{tree-search}
method. Fig. \ref{fig:SM-tree-search.} illustrates the tree-search
concept for the SM demodulation with $M=2$, $N_{t}=2$, and $N_{r}=3$.
In the SM tree-search method, each possible combination of $\mathbf{x}_{j}$
in (\ref{eq: x_est_ML_index}) is represented by a tree-search branch
whose length is $N_{r}$ tree-search nodes (or levels). Each node
is an accumulation of the previous EDs in the same branch, which can
be represented as

\begin{equation}
d_{i,j}=\sum_{n=1}^{i}\left|y_{n}-x_{n,j}\right|^{2},\,\,\,\,\,\,\,\,\,i=1,\ldots,N_{r},\label{eq: d_i,j}
\end{equation}

\noindent where $d_{i,j}$ is the node metric at the $i$-th level
of the $j$-th branch. Hence, (\ref{eq: x_est_ML_index}) can be rewritten
as

\begin{equation}
\hat{j}_{\text{ML}}=\underset{j=1,\ldots,MN_{t}}{\text{arg}\,\text{min}}\left\{ d_{N_{r},j}\right\} .\label{eq: x_est_ML (ED)}
\end{equation}

\noindent Thus, the ML solution for the estimated transmitted combination
is denoted by $\hat{\mathbf{x}}_{\text{ML}}$ and given as

\begin{equation}
\hat{\mathbf{x}}_{\text{ML}}=\mathbf{x}_{\hat{j}_{\text{ML}}}.\label{eq: x_est_ML}
\end{equation}

\noindent The total number of nodes for the SM tree-search is $MN_{t}N_{r}$,
which is $12$ in the example of Fig. \ref{fig:SM-tree-search.}.
To estimate the transmitted combination using the ML detection algorithm,
the DSP unit exhaustively visits all nodes, which can be problematic
for increasing values of $M$, $N_{t}$ and $N_{r}$. Thus, reducing
the decoding complexity has paramount importance for real-time applications.

\section{\label{sec:Minimum-Distance-Maximum-Length-}Minimum-Distance of
Maximum-Length Algorithm}

Unlike the existent SD algorithms in the literature, the proposed
m-M algorithm performs only one node expansion at a time; the expanded
node is chosen to be of minimum ED across all branches. The proposed
algorithm jumps from one branch to another according to where the
minimum ED is, and stops if the minimum ED occurs at the end of a
fully expanded branch (i.e., maximum length).

For mathematical formulation, assume that $\mathbf{v}=\left[v_{1}\ldots v_{MN_{t}}\right]\in\mathbb{R}^{1\times MN_{t}}$
denote the vector of visited nodes, where $v_{j}$ takes integer values
from $1$ up to $N_{r}$ and represents the number of nodes already
visited of the $j$-th branch for $j=1,\ldots,MN_{t}$. Also, let
$\mathbf{d}=\left[d_{v_{1},1}\ldots d_{v_{MN_{t}},MN_{t}}\right]\in\mathbb{R}^{1\times MN_{t}}$
denote the ED vector, where $d_{v_{j},j}$ is given by (\ref{eq: d_i,j})
by setting $i=v_{j}$ (i.e., $d_{v_{j},j}=\sum_{n=1}^{v_{j}}\left|y_{n}-x_{n,j}\right|^{2}$,
where $d_{v_{j},j}$ represents the ED (node metric) of the $v_{j}$-th
level for the $j$-th branch).

\noindent Algorithm \ref{alg: min-max algorithm} summarizes the proposed
m-M algorithm that is explained as follows:

\textbf{Step 1:} Initialize all elements of $\mathbf{v}$ to unity
(i.e., $v_{j}=1\,\,\forall j$), and then calculate each element of
the vector $\mathbf{d}$ from (\ref{eq: d_i,j}) accordingly. It should
be noted that the elements of $\mathbf{d}$ in this step represent
the first ED of all branches (i.e., $\mathbf{d}=[d_{1,1}\ldots d_{1,MN_{t}}]$).

\textbf{Step 2:} Determine the argument of the minimum element of
$\mathbf{d}$ as

\begin{equation}
j_{\text{min}}=\underset{j=1,\ldots,MN_{t}}{\text{arg}\,\text{min}}\{d_{v_{j},j}\}.\label{eq: j_min}
\end{equation}

\textbf{Step 3:} Increase the $j_{\text{min}}$-th element of $\mathbf{v}$
by one 

\begin{equation}
v_{j_{\text{min}}}\rightarrow v_{j_{\text{min}}}+1.\label{eq: increment v}
\end{equation}

\noindent Note that this step ensures that the algorithm makes a single
expansion to the minimum ED, which leads to the increase of the corresponding
element of the vector $\mathbf{v}$ by one. The maximum value of $v_{j}\,\,\forall j$
that can be reached is $N_{r}$; therefore, we can define $\mathbf{j}_{\text{max}}$
as the set of indices whose values reached $N_{r}$, as

\begin{equation}
\mathbf{j}_{\text{max}}=\text{find}\left(\mathbf{v}=N_{r}\right),\label{eq:j_max}
\end{equation}

\noindent where $\text{find}\left(\mathbf{v}=N_{r}\right)$ returns
the indices of the elements of $\mathbf{v}$ that are equal to $N_{r}$.
At the beginning, $\mathbf{j}_{\text{max}}$ is buffered as an empty
set, and is updated when at least one branch is fully expanded.

\textbf{Step 4:} Update the $j_{\text{min}}$-th element of $\mathbf{d}$
by calculating the new $d_{v_{j_{\text{min}}},j_{\text{min}}}$ from
(\ref{eq: d_i,j}) based on $v_{j_{\text{min}}}$ calculated from
\textbf{Step 3}.

\textbf{Step 5:} Find the new $j_{\text{min}}$ from (\ref{eq: j_min})
as in \textbf{Step 2}, and then check whether the following condition
is true or not:

\noindent 
\begin{equation}
j_{\text{min}}\in\mathbf{j}_{\text{max}}.\label{eq: opt condition}
\end{equation}

\noindent If $j_{\text{min}}\notin\mathbf{j}_{\text{max}}$, then
go back to \textbf{Step 2}. Otherwise, find the index of the estimated
transmitted combination as 

\begin{equation}
\hat{j}_{\text{m-M}}=\underset{j\in\mathbf{j}_{\text{max}}}{\text{arg}\,\text{min}}\{d_{v_{j},j}\},\label{eq: index_x_est_m_M}
\end{equation}

\noindent where $\hat{j}_{\text{m-M}}$ denotes the index of the estimated
transmitted combination from the m-M algorithm. Note that in case
of $v_{j}=N_{r}\,\,\forall j$ in (\ref{eq: index_x_est_m_M}), the
ML version in (\ref{eq: x_est_ML (ED)}) is obtained. The estimated
transmitted combination from m-M algorithm, $\hat{\mathbf{x}}_{\text{m-M}}$,
is

\begin{equation}
\hat{\mathbf{x}}_{\text{m-M}}=\mathbf{x}_{\hat{j}_{\text{m-M}}}.\label{eq:x_m_M}
\end{equation}

\noindent Note that the condition in (\ref{eq: opt condition}) is
called the optimality condition, and guarantees that the ML solution
will not be missed before stopping the m-M algorithm (i.e., $\hat{\mathbf{x}}_{\text{m-M}}=\hat{\mathbf{x}}_{\text{ML}}$). 

\begin{algorithm}[t]
\begin{itemize}
\item \textbf{\small{}Initialize} $\mathbf{v}=\left[1\,\,1\,\ldots\,1\right]\in\mathbb{R}^{1\times MN_{t}}$,
$j_{\text{max}}=0$.
\item \textbf{\small{}Compute }{\small{}the elements of $\mathbf{d}=[d_{1,1}\ldots d_{1,MN_{t}}]$,
where $d_{1,j}=\left|y_{1}-x_{1,j}\right|^{2}$ and $j=1,\ldots,MN_{t}$.}{\small\par}
\item \textbf{Reserve} an empty vector $\mathbf{j}_{\text{max}}=[.]$ as
a buffer.
\end{itemize}
{\small{}~~~~~1: $\text{\textbf{while}}\,\,\,n\leq N_{r}MN_{t}$
}\textbf{\small{}~do}{\small\par}

{\small{}~~~~~2:}\textbf{\small{} ~~~Find}{\small{} the index
$j_{\text{min}}=\underset{j=1,\ldots,MN_{t}}{\text{arg}\,\text{min}}\{d_{v_{j},j}\}$.}{\small\par}

{\small{}~~~~~3: ~~~}\textbf{\small{}if}{\small{} $\mathbf{j}_{\text{max}}$
is NOT empty}{\small\par}

{\small{}~~~~~4:}\textbf{\small{} ~~~~~if}{\small{} $j_{\text{min}}\in\mathbf{j}_{\text{max}}$ }{\small\par}

{\small{}~~~~~5:}\textbf{\small{} ~~~~~~~~go to}{\small{}
line 14.}{\small\par}

{\small{}~~~~~6:}\textbf{\small{} ~~~~~else}{\small\par}

{\small{}~~~~~7:}\textbf{\small{} ~~~~~~~~go to}{\small{}
line 10.}{\small\par}

{\small{}~~~~~8:}\textbf{\small{} ~~~~~end if}{\small{} }{\small\par}

{\small{}~~~~~9: ~~~}\textbf{\small{}end if}{\small\par}

{\small{}~~~~10: }\textbf{\small{}~~~Set}{\small{} $v_{j_{\text{min}}}\rightarrow v_{j_{\text{min}}}+1$,
then }\textbf{\small{}Update}{\small{} $\mathbf{v}$.}{\small\par}

{\small{}~~~~11: }\textbf{\small{}~~~Update}{\small{} }the
$j_{\text{min}}$-th element of $\mathbf{d}${\small{} as:}{\small\par}

\hspace{24.5mm}$d_{v_{j_{\text{min}}},j_{\text{min}}}\rightarrow d_{v_{j_{\text{min}}},j_{\text{min}}}+\left|y_{v_{j_{\text{min}}}}-x_{v_{j_{\text{min}}},j_{\text{min}}}\right|^{2}${\small{}.}{\small\par}

{\small{}~~~~12: }\textbf{\small{}~~~Update}{\small{} $\mathbf{j}_{\text{max}}$
based on $\mathbf{j}_{\text{max}}=\text{find}\left(\mathbf{v}=N_{r}\right)$.}{\small\par}

{\small{}~~~~13: }\textbf{\small{}~~~Set} $n\rightarrow n+1$.

{\small{}~~~~14: }\textbf{\small{}end while}{\small\par}
\begin{itemize}
\item \textbf{\small{}Estimate}{\small{} }$\hat{\mathbf{x}}_{\text{m-M}}$
{\small{}from $\hat{\mathbf{x}}_{\text{m-M}}=\mathbf{x}_{j_{\text{min}}}$.}{\small\par}
\end{itemize}
\caption{\label{alg: min-max algorithm}Pseudo-code of the proposed m-M algorithm.}
\end{algorithm}

\begin{figure*}
\begin{centering}
\includegraphics[width=7.5cm,height=6cm]{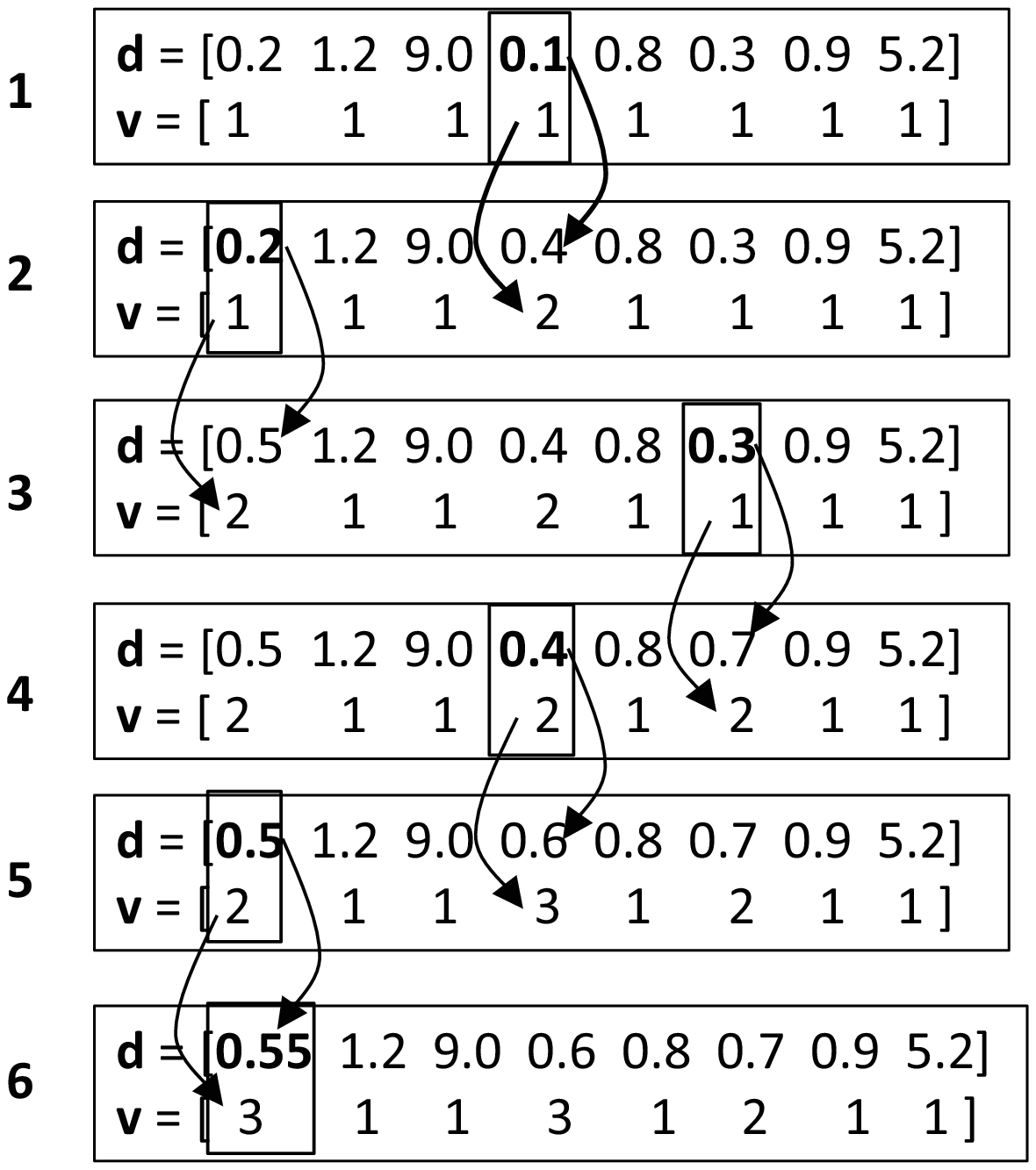}\includegraphics[width=7.5cm,height=6cm]{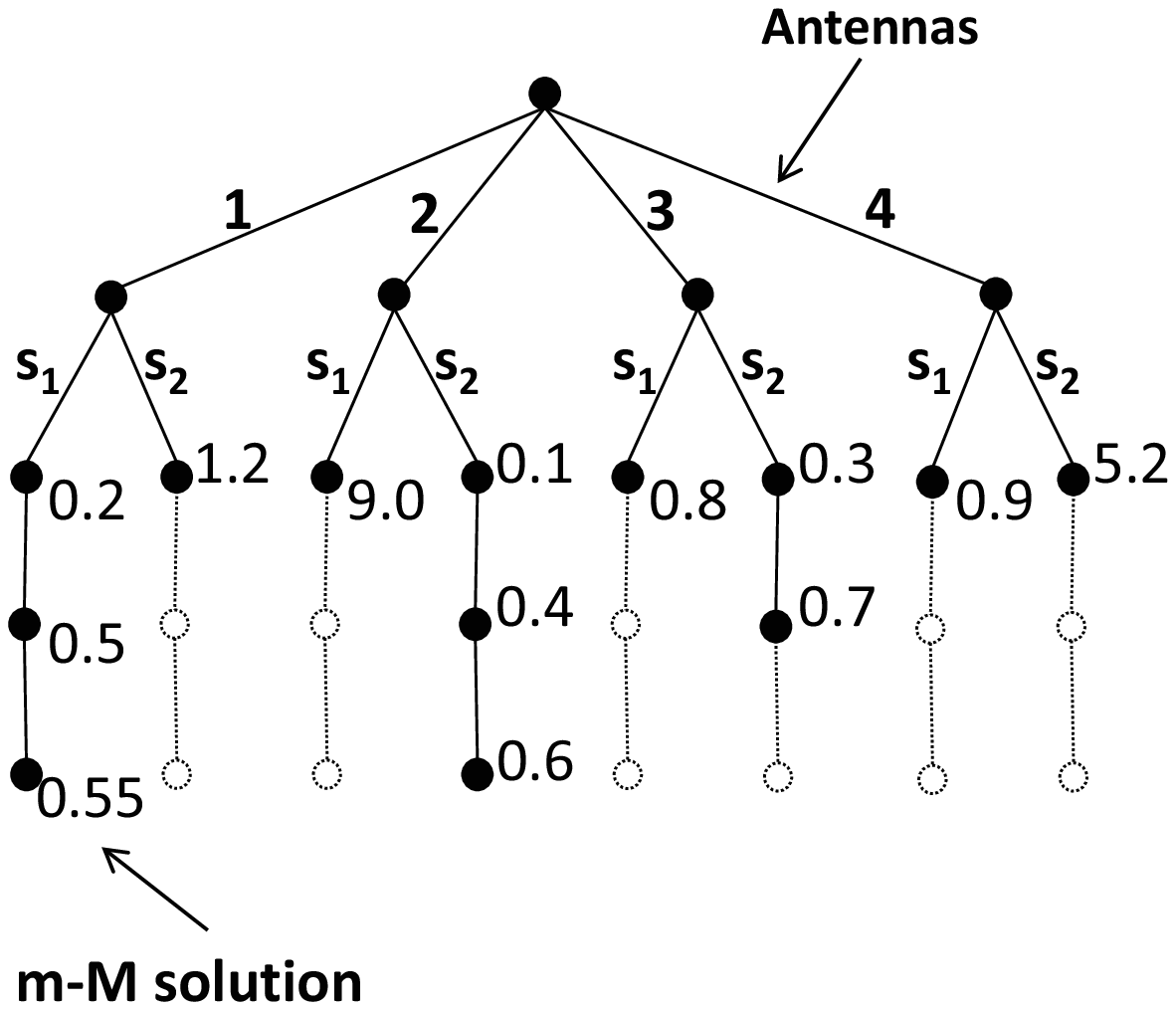}
\par\end{centering}
\caption{{\small{}\label{fig:Numerical-example-of}A numerical example for
the m-M algorithm ($3\times4$ MIMO-SM system and $M=2$).}}
\end{figure*}

Fig. \ref{fig:Numerical-example-of} illustrates a numerical example
for the proposed m-M algorithm. Consider a $3\times4$ MIMO system
with $M=2$. Thus, we have $8$ branches with $3$ nodes/levels length.
First, the m-M algorithm initializes $\mathbf{v}$ by all-ones, and
calculates the the first ED of each branch. The m-M algorithm finds
the minimum ED of $\mathbf{d}$, which is $0.1$ in our example. This
ED corresponds to the $4$-th branch ($j_{\text{min}}=4$); thus,
the m-M algorithm expands this node after increasing the $4$-th element
of $\mathbf{v}$ by one (i.e., $v_{j_{\text{min}}}=v_{4}=2$ and $d_{2,4}=0.4$).
In the second iteration, the m-M algorithm finds the new minimum ED
in $\mathbf{d}$ (i.e., 0.2), which is placed in the first branch
($j_{\text{min}}=1$). Then, the first element of $\mathbf{v}$ is
updated to be $2$ and the first element of $\mathbf{d}$ is updated
accordingly (i.e., $v_{1}=2$ and $d_{2,1}=0.5$). The algorithm jumps
from one branch to another according to the location of the minimum
ED across all branches, as illustrated in iterations 3, 4, and 5.
Note that the m-M algorithm detects one element of $\mathbf{v}$ reaches
$N_{r}$ (i.e., full expansion for that branch) from iteration 5,
which is the $4$-th branch. According to (\ref{eq: opt condition}),
the algorithm has to check if the new minimum ED comes at a fully
expanded branch or not before deciding to stop. In our example, the
algorithm will not stop at iteration 5 because there is a minimum
ED at the first branch (i.e., 0.5). Therefore, the algorithm makes
a single expansion to the first branch after updating the first element
of $\mathbf{v}$ (i.e., $v_{1}=3$ and $d_{3,1}=0.55$); and then,
it checks the place of the minimum ED once more. In this example,
the iteration 6 shows that the minimum ED (i.e., 0.55) comes at the
end of a fully expanded branch, which corresponds to the first branch
(i.e., $\hat{j}_{\text{m-M}}=1$). Thus, the m-M algorithm stops and
declares that the estimated transmitted combination is the first one
(i.e., the first symbol was transmitted from the first antenna).

\section{\label{sec:Complexity-Analysis}Complexity Analysis}

In this paper, we consider the number of visited nodes inside the
tree-search as the complexity indicator. Since $\mathbf{v}$ represents
the visited nodes for each branch, the summation of its elements at
the final iteration gives the total complexity of the m-M algorithm
in terms of the number of visited nodes. Consider the complexity of
the m-M algorithm denoted by $C_{\text{m-M}}=\text{sum}\left\{ \mathbf{v}^{\text{f}}\right\} $,
where $\mathbf{v}^{\text{f}}$ is the vector $\mathbf{v}$ at the
final iteration. Since the elements of $\mathbf{v}^{\text{f}}$ are
random variables (r.v.'s), $C_{\text{m-M}}$ is an r.v. as well. In
this section, we provide a tight expression for the expected complexity
of the proposed m-M algorithm in the case of perfect CSIR, as well
as imperfect CSIR.

The average complexity of the m-M algorithm $C_{\text{m-M}}$ can
be expressed as

\begin{equation}
C_{\text{m-M}}=\mathbb{E}\left\{ \text{sum}\left\{ \mathbf{v}^{\text{f}}\right\} \right\} .\label{eq: C_m_M}
\end{equation}

\noindent Although the m-M algorithm is a breadth-first search algorithm,
its expected complexity is equivalent to that of a depth-first SD
algorithm with pruned radius, $R_{\text{m-M}}$, equal to the minimum
ED of vector $\mathbf{d}$ at the final iteration (i.e., $0.55$ in
the example illustrated in Fig. \ref{fig:Numerical-example-of}).
Therefore, $R_{\text{m-M}}$ can be written as

\begin{equation}
R_{\text{m-M}}=d_{N_{r},\hat{j}_{\text{m-M}}}=\sum_{n=1}^{N_{r}}\left|y_{n}-x_{n,\hat{j}_{\text{m-M}}}\right|^{2}=\left\Vert \mathbf{y}-\hat{\mathbf{x}}_{\text{m-M}}\right\Vert ^{2},\label{eq: R_m_M}
\end{equation}

\noindent where $\hat{j}_{\text{m-M}}$ given from (\ref{eq: index_x_est_m_M})
and $\hat{\mathbf{x}}_{\text{m-M}}$ are given in (\ref{eq: j_min})
and (\ref{eq:x_m_M}), respectively. For simplicity, we consider $\hat{\mathbf{x}}_{\text{m-M}}\rightarrow\mathbf{x}_{t}$;
this assumption most likely holds particularly in high signal-to-noise
ratio (SNR) ($\hat{\mathbf{x}}_{\text{m-M}}=\hat{\mathbf{x}}_{\text{ML}}$
since the m-M algorithm guarantees the ML solution). Thus, substituting
(\ref{eq: y}) and this assumption in (\ref{eq: R_m_M}) yields 

\begin{equation}
R_{\text{m-M}}=\left\Vert \mathbf{w}\right\Vert ^{2}.\label{eq:R_m_M assumption}
\end{equation}

\noindent It should be noted that the pruned radius in (\ref{eq: R_m_M})
is considered the optimum threshold that can be used in the SD-based
algorithms. Since the decoding complexity of the proposed m-M algorithm
is equivalent to that of a depth-first algorithm using the optimum
pruned radius in (\ref{eq: R_m_M}), the proposed algorithm provides
a better complexity than the optimum BER algorithms in the literature.

Now, we can write $C_{\text{m-M}}$ in (\ref{eq: C_m_M}) as {[}\ref{A.-Younis,-R.GLOBOCOM}{]},
{[}\ref{I.-Al-Nahhal,-O. QSM}{]}

\begin{equation}
C_{\text{m-M}}\approx MN_{t}+\sum_{j=1}^{MN_{t}}\sum_{i=1}^{N_{r}}\mathbb{P}\text{r}\left(d_{i,j}\leq R_{\text{m-M}}\left|\mathbf{x}_{t},\mathbf{H},\sigma_{n}^{2},R_{\text{m-M}}\right.\right).\label{eq: C_m_M _expected}
\end{equation}

\noindent It is worth noting that (\ref{eq: C_m_M _expected}) is
the generic form of the expected complexity, and its closed-form solution
depends on the algorithm itself. Note that (\ref{eq: C_m_M _expected})
finds the probability of $d_{i,j}$ being visited when the SD radius
is $R_{\text{m-M}}$ (the node is considered to be visited if $d_{i,j}\leq R_{\text{m-M}}$
and vice versa). Ideally, $\mathbb{P}\text{r}\left(d_{i,j}\leq R_{\text{m-M}}\right)$
under the conditions previously given should be zero or one. The correction
factor $MN_{t}$ in (\ref{eq: C_m_M _expected}) is needed since the
$\mathbb{P}\text{r}\left(d_{i,j}\leq R_{\text{m-M}}\right)$ misses
almost $MN_{t}$ nodes at the final iteration.

\subsection{Perfect Channel State Information at the Receiver}

To find the closed form expression of the right-hand-side of (\ref{eq: C_m_M _expected}),
the conditional probability distribution of $d_{i,j}$ should be determined
first. From (\ref{eq: y}) and (\ref{eq: d_i,j}), we can rewrite
(\ref{eq: d_i,j}) in terms of the real and imaginary components as\\

$d_{i,j}=\sum_{n=1}^{i}\left|\left(w_{n}^{\Re}+x_{n,t}^{\Re}-x_{n,j}^{\Re}\right)+\jmath\left(w_{n}^{\Im}+x_{n,t}^{\Im}-x_{n,j}^{\Im}\right)\right|^{2}$

\vspace{-2mm}

\begin{equation}
=\sum_{n=1}^{i}\left(\mathcal{R}_{n}^{2}+\mathcal{I}_{n}^{2}\right),\hspace{10mm}\label{eq:d_i_j_real_imag}
\end{equation}

\noindent where $\mathcal{R}_{n}=w_{n}^{\Re}+x_{n,t}^{\Re}-x_{n,j}^{\Re}$
and $\mathcal{I}_{n}=w_{n}^{\Im}+x_{n,t}^{\Im}-x_{n,j}^{\Im}$ are
Gaussian distributed with variances $\sigma_{n}^{2}/2$, and means
$\left(x_{n,t}^{\Re}-x_{n,j}^{\Re}\right)$ and $\left(x_{n,t}^{\Im}-x_{n,j}^{\Im}\right)$,
respectively. Consequently, $d_{i,j}$ is a non-central chi-squared
r.v. with $2i$ degrees of freedom and non-centrality parameter $\gamma_{i,j}^{2}$
given by {[}\ref{J.-Proakis,-Digital}, (Ch. 2){]}

\begin{equation}
\gamma_{i,j}^{2}=\sum_{n=1}^{i}\left[\left(x_{n,t}^{\Re}-x_{n,j}^{\Re}\right)^{2}+\left(x_{n,t}^{\Im}-x_{n,j}^{\Im}\right)^{2}\right].\label{eq: non-cenral_parameter}
\end{equation}

\noindent The probability distribution function (pdf) of $d_{i,j}$
for $d_{i,j}\geq0$ is calculated as {[}\ref{J.-Proakis,-Digital},
(Ch. 2){]}\\

$f_{d_{i,j}}(d_{i,j})=\frac{1}{\sigma_{n}^{2}}\left(\frac{d_{i,j}}{\gamma_{i,j}^{2}}\right)^{(i-1)/2}$

\vspace{-2mm}

\begin{equation}
\times\,\text{exp}\left(-\frac{\gamma_{i,j}^{2}+d_{i,j}}{\sigma_{n}^{2}}\right)\,I_{i-1}\left(\frac{\sqrt{d_{i,j}\,\gamma_{i,j}^{2}}}{\sigma_{n}^{2}/2}\right),\label{eq: d_i_j pdf}
\end{equation}

\noindent where $I_{i-1}\left(\centerdot\right)$ is the first kind
modified Bessel function of order $(i-1)$. Since $d_{i,j}$ has an
even degrees of freedom, the closed form expression of the cumulative
distribution function (cdf) for (\ref{eq: d_i_j pdf}) is given as
{[}\ref{J.-Proakis,-Digital}, (Ch. 2){]}\\

$\mathbb{P}\text{r}\left(d_{i,j}\leq R_{\text{m-M}}\left|\mathbf{x}_{t},\mathbf{H},\sigma_{n}^{2},R_{\text{m-M}}\right.\right)=$

\vspace{-2mm}

\begin{equation}
\hspace{35mm}1-Q_{i}\left(\frac{\gamma_{i,j}}{\sigma_{n}/\sqrt{2}}\,,\,\frac{\sqrt{R_{\text{m-M}}}}{\sigma_{n}/\sqrt{2}}\right),\label{eq:CDF_PCSI}
\end{equation}

\noindent where $Q_{i}(\centerdot,\centerdot)$ denotes the generalized
Marcum function of order $i$.

To remove the dependency of (\ref{eq:CDF_PCSI}) on the instantaneous
value of $R_{\text{m-M}}$, an expectation over the pdf of $R_{\text{m-M}}$
should be calculated. (\ref{eq:R_m_M assumption}) can be written
in terms of its real and imaginary components as

\begin{equation}
R_{\text{m-M}}=\sum_{n=1}^{N_{r}}\left[\left(w_{n}^{\Re}\right)^{2}+\left(w_{n}^{\Im}\right)^{2}\right].\label{eq: R_m_M real =000026 imag}
\end{equation}

\noindent Therefore, $R_{\text{m-M}}$ is a central chi-square r.v.
with $2N_{r}$ degrees of freedom and its pdf, $f_{R_{\text{m-M}}}(R_{\text{m-M}})$,
is {[}\ref{J.-Proakis,-Digital}, (Ch. 2){]}

\begin{equation}
f_{R_{\text{m-M}}}(R_{\text{m-M}})=\frac{\left(R_{\text{m-M}}\right)^{N_{r}-1}}{\sigma_{n}^{2N_{r}}\left(N_{r}-1\right)!}\,\text{exp}\left(\frac{-R_{\text{m-M}}}{\sigma_{n}^{2}}\right).\label{eq:R_m_M pdf}
\end{equation}

From (\ref{eq:CDF_PCSI}) and (\ref{eq:R_m_M pdf}), the expected
value of (\ref{eq:CDF_PCSI}) over the pdf of $R_{\text{m-M}}$ can
be written as\\

$\mathbb{P}\text{r}\left(d_{i,j}\leq R_{\text{m-M}}\left|\mathbf{x}_{t},\mathbf{H},\sigma_{n}^{2}\right.\right)=$

\vspace{-2mm}

\begin{equation}
\int_{0}^{\infty}\left[1-Q_{i}\left(\frac{\gamma_{i,j}}{\sigma_{n}/\sqrt{2}}\,,\,\frac{\sqrt{R_{\text{m-M}}}}{\sigma_{n}/\sqrt{2}}\right)\right]f_{R_{\text{m-M}}}(R_{\text{m-M}})\,\,\,\,\text{d}R_{\text{m-M}}.\label{eq:CDF independent}
\end{equation}

\noindent The closed form solution of the integration in (\ref{eq:CDF independent})
can be found in {[}\ref{P.-C.-Sofotasios, solution to integ}{]},
and then, the complexity in (\ref{eq: C_m_M _expected}) is expressed
as 

\begin{equation}
\begin{gathered}C_{\text{m-M}}\approx MN_{t}+\sum_{j=1}^{MN_{t}}\sum_{i=1}^{N_{r}}\left[1-\left[1-\frac{\text{exp}\left(-\gamma_{i,j}^{2}/\sigma_{n}^{2}\right)}{2^{N_{r}}}\right.\right.\\
\times\left[\Phi_{1}\left(N_{r},1,1;\frac{1}{2},\frac{\gamma_{i,j}^{2}}{2\sigma_{n}^{2}}\right)\hspace{12.5mm}\right.\\
\left.\left.\left.\hspace{14mm}-\sum_{k=1}^{i-1}\frac{\left(N_{r}\right)_{k}}{2^{k}\,k!}\,_{1}F_{1}\left(N_{r}+k;k+1;\frac{\gamma_{i,j}^{2}}{2\sigma_{n}^{2}}\right)\right]\right]\right],
\end{gathered}
\label{eq:C_m_M PCSI}
\end{equation}

\noindent where $\left(N_{r}\right)_{k}$ denotes the Pochhammer symbol,
$\Phi_{1}$ is the Humbert hypergeometric function of the first kind,
and $_{1}F_{1}$ denotes the Kummer hypergeometric function {[}\ref{Y.-A.-Brychkov, special functions}{]}.

\subsection{Imperfect Channel State Information at the Receiver}

In this subsection, the complexity of the proposed m-M algorithm in
(\ref{eq: C_m_M _expected}) is assessed in the presence of imperfect
CSIR. To the best of the authors' knowledge, in case of imperfect
CSIR, the expected complexity is not analyzed in the literature. Assume
that there is an error between the estimated channel coefficient at
the receiver side and the actual channel coefficient, which is denoted
by $\mathbf{e}\sim\mathcal{C}\mathcal{N}\left(0,\sigma_{e}^{2}\right)$,
where $\sigma_{e}^{2}$ is the variance of the error in the channel
estimation. Thus, the estimated channel entry becomes $\hat{\mathbf{h}}=\mathbf{h}+\mathbf{e}$
and the combination element in (\ref{eq: d_i,j}) becomes $x_{n,j}+\hat{e}_{n,j}$,
where $\hat{e}_{n,j}=e_{n}s_{j}$, with $s_{j}$ as the QAM symbol
in $j$-th combination with energy of $\left|s_{j}\right|^{2}$ and
$e_{n}$ as the $n$-th element of vector $\mathbf{e}$. In this case,
for least square solution of (\ref{eq: d_i,j}), $\hat{\mathbf{h}}\sim\mathcal{C}\mathcal{N}\left(0,1+\sigma_{e}^{2}\right)$
depends on $\mathbf{h}$ with a correlation coefficient of $\rho=1/\sqrt{1+\sigma_{e}^{2}}$
{[}\ref{J.-Wu-and}{]}-{[}\ref{E.-Ba=00015Far,-=0000DC. imperfect csi}{]},
{[}\ref{A.-Leon-Garcia,-Probability,}, (p. 282){]}; the conditional
variance of the elements of the noisy received vector, $\zeta_{j}^{2}$,
is given by {[}\ref{E.-Ba=00015Far,-=0000DC. imperfect csi}{]}, {[}\ref{V.-Tarokh,-A.}{]}

\begin{equation}
\zeta_{j}^{2}=\text{Var}\left(y|\hat{\mathbf{h}}\right)=\sigma_{n}^{2}+\left(1-\rho^{2}\right)\left|s_{j}\right|^{2}.\label{eq: VAR_j}
\end{equation}

\noindent It should be noted that the $\sigma_{e}^{2}$ may be considered
as fixed or variable when SNR changes. In theory, the error in channel
estimation decreases as the SNR increases {[}\ref{M.-Biguesh-and}{]},
{[}\ref{M.-Biguesh,-and}{]}; therefore, we can consider $\sigma_{e}^{2}=1/\text{snr}$
in case of variable $\sigma_{e}^{2}$ where snr denotes the signal-to-noise
ratio in linear scale (i.e., SNR = $10\text{lo\ensuremath{g_{10}}}\left(\text{snr}\right)$). 

$d_{i,j}$ in (\ref{eq:d_i_j_real_imag}) in the case of imperfect-CSIR
is denoted by $\hat{d}_{i,j}$ and given as\\

$\hat{d}_{i,j}=\sum_{n=1}^{i}\left|\left(w_{n}^{\Re}-\hat{e}_{n,j}^{\Re}+x_{n,t}^{\Re}-x_{n,j}^{\Re}\right)\right.$

\vspace{-2mm}

\begin{equation}
\left.+\jmath\left(w_{n}^{\Im}-\hat{e}_{n,j}^{\Im}+x_{n,t}^{\Im}-x_{n,j}^{\Im}\right)\right|^{2}=\sum_{n=1}^{i}\left(\mathcal{\hat{R}}_{n}^{2}+\hat{\mathcal{I}}_{n}^{2}\right),\label{eq:d_i_j_real_imag Imperfect CSI}
\end{equation}

\noindent where $\mathcal{\hat{R}}_{n}=w_{n}^{\Re}-\hat{e}_{n,j}^{\Re}+x_{n,t}^{\Re}-x_{n,j}^{\Re}$
and $\hat{\mathcal{I}}_{n}=w_{n}^{\Im}-\hat{e}_{n,j}^{\Im}+x_{n,t}^{\Im}-x_{n,j}^{\Im}$
are Gaussian distributed with variances $\zeta_{j}^{2}/2$, and means
$\left(x_{n,t}^{\Re}-x_{n,j}^{\Re}\right)$ and $\left(x_{n,t}^{\Im}-x_{n,j}^{\Im}\right)$,
respectively. Consequently, $\hat{d}_{i,j}$ is a non-central chi-squared
r.v. with $2i$ degrees of freedom and non-centrality parameter $\gamma_{i,j}^{2}$
given by (\ref{eq: non-cenral_parameter}), and its pdf for $\hat{d}_{i,j}\geq0$
becomes {[}\ref{J.-Proakis,-Digital}, (Ch. 2){]}\\

$f_{\hat{d}_{i,j}}(\hat{d}_{i,j})=\frac{1}{\zeta_{j}^{2}}\left(\frac{\hat{d}_{i,j}}{\gamma_{i,j}^{2}}\right)^{(i-1)/2}$

\vspace{-2mm}

\begin{equation}
\times\,\text{exp}\left(-\frac{\gamma_{i,j}^{2}+\hat{d}_{i,j}}{\zeta_{j}^{2}}\right)\,I_{i-1}\left(\frac{\sqrt{\hat{d}_{i,j}\,\gamma_{i,j}^{2}}}{\zeta_{j}^{2}/2}\right).\label{eq: d_i_j pdf-1}
\end{equation}

Therefore, (\ref{eq:CDF_PCSI}) becomes\\

$\mathbb{P}\text{r}\left(\hat{d}_{i,j}\leq\hat{R}_{\text{m-M}}\left|\mathbf{x}_{t},\mathbf{H},\sigma_{n}^{2},\sigma_{e}^{2},\hat{R}_{\text{m-M}}\right.\right)=$

\vspace{-2mm}

\begin{equation}
\hspace{35mm}1-Q_{i}\left(\frac{\gamma_{i,j}}{\zeta_{j}/\sqrt{2}}\,,\,\frac{\sqrt{\hat{R}_{\text{m-M}}}}{\zeta_{j}/\sqrt{2}}\right),\label{eq:CDF_Imperfect CSI}
\end{equation}

\noindent where $\hat{R}_{\text{m-M}}$ denotes the threshold of the
m-M algorithm in the case of imperfect CSIR. It should be noted that
for the case of imperfect CSIR, the threshold in (\ref{eq: R_m_M real =000026 imag})
becomes

\begin{equation}
\hat{R}_{\text{m-M}}=\sum_{n=1}^{N_{r}}\left[\left(w_{n}^{\Re}-\hat{e}_{n,t}^{\Re}\right)^{2}+\left(w_{n}^{\Im}-\hat{e}_{n,t}^{\Im}\right)^{2}\right],\label{eq: R_m_M real =000026 imag Imperfect CSI}
\end{equation}

\noindent where $\left(w_{n}^{\Re}-\hat{e}_{n,t}^{\Re}\right)$ and
$\left(w_{n}^{\Im}-\hat{e}_{n,t}^{\Im}\right)$ are Gaussian distributed
with zero-mean and variance of $\zeta_{t}^{2}/2$, where

\noindent 
\begin{equation}
\zeta_{t}^{2}=\sigma_{n}^{2}+\left(1-\rho^{2}\right)\left|s_{t}\right|^{2},\label{eq: VAR_t}
\end{equation}

\noindent with $s_{t}$ as the transmitted QAM symbol with energy
$\left|s_{t}\right|^{2}$.

\noindent Consequently, $\hat{R}_{\text{m-M}}\geq0$ is a central
chi-squared distributed r.v. with $2N_{r}$ degrees of freedom and
its pdf is given by {[}\ref{J.-Proakis,-Digital}, (Ch. 2){]}

\begin{equation}
f_{\hat{R}_{\text{m-M}}}(\hat{R}_{\text{m-M}})=\frac{\left(\hat{R}_{\text{m-M}}\right)^{N_{r}-1}}{\zeta_{t}^{2N_{r}}\left(N_{r}-1\right)!}\hspace{1.5mm}\text{exp}\left(\frac{-\hat{R}_{\text{m-M}}}{\zeta_{t}^{2}}\right).\label{eq:R_m_M pdf imperfect CSI}
\end{equation}

From (\ref{eq:CDF_Imperfect CSI}) and (\ref{eq:R_m_M pdf imperfect CSI}),
the expected value of (\ref{eq:CDF_Imperfect CSI}) over the pdf of
$\hat{R}_{\text{m-M}}$ can be written as\\

$\mathbb{P}\text{r}\left(\hat{d}_{i,j}\leq\hat{R}_{\text{m-M}}\left|\mathbf{x}_{t},\mathbf{H},\sigma_{n}^{2},\sigma_{e}^{2}\right.\right)=$

\vspace{-2mm}

\begin{equation}
\int_{0}^{\infty}\left[1-Q_{i}\left(\frac{\gamma_{i,j}}{\zeta_{j}/\sqrt{2}}\,,\,\frac{\sqrt{\hat{R}_{\text{m-M}}}}{\zeta_{j}/\sqrt{2}}\right)\right]f_{\hat{R}_{\text{m-M}}}(\hat{R}_{\text{m-M}})\,\,\,\,\text{d}\hat{R}_{\text{m-M}}.\label{eq:CDF independent imperfect CSI}
\end{equation}

\noindent The closed form of the integration in (\ref{eq:CDF independent imperfect CSI})
can be found in {[}\ref{P.-C.-Sofotasios, solution to integ}{]},
and then, the complexity in (\ref{eq: C_m_M _expected}) is obtained
as

\begin{equation}
\begin{gathered}\hat{C}_{\text{m-M}}\approx MN_{t}+\sum_{j=1}^{MN_{t}}\sum_{i=1}^{N_{r}}\left[1-\left[1-\frac{\zeta_{j}^{2N_{r}}\,\text{exp}\left(-\gamma_{i,j}^{2}/\zeta_{j}^{2}\right)}{\left(\zeta_{j}^{2}+\zeta_{t}^{2}\right)^{N_{r}}}\right.\right.\\
\times\left[\Phi_{1}\left(N_{r},1,1;\frac{\zeta_{t}^{2}}{\zeta_{j}^{2}+\zeta_{t}^{2}},\frac{\gamma_{i,j}^{2}\,\zeta_{t}^{2}}{\zeta_{j}^{2}\left(\zeta_{j}^{2}+\zeta_{t}^{2}\right)}\right)\hspace{12.5mm}\right.\\
\left.\left.\left.\hspace{5mm}-\sum_{k=1}^{i-1}\frac{\left(N_{r}\right)_{k}}{k!}\,_{1}F_{1}\left(N_{r}+k;k+1;\frac{\gamma_{i,j}^{2}\,\zeta_{t}^{2}}{\zeta_{j}^{2}\left(\zeta_{j}^{2}+\zeta_{t}^{2}\right)}\right)\right]\right]\right].
\end{gathered}
\label{eq:C_m_M imperfect CSI}
\end{equation}

\section{\label{sec:Optimality-of-BER-Performance}Optimality of BER Performance}

In this section, we discuss the BER performance optimality of the
proposed m-M algorithm based on the condition in (\ref{eq: opt condition}).
The effect of omitting this condition on the proposed m-M\textcolor{blue}{{}
}algorithm is also studied. We define an indicator for the BER performance
optimality as the number of times the proposed m-M algorithm misses
the ML solution, referred to as the number of misses (NoM). In other
words, the BER of the m-M algorithm will be the same as the ML BER
if the NoM equals zero and vice versa. It should be noted that NoM
is an r.v. that depends on the SNR and $\sigma_{e}^{2}$.

Let us invoke the general expression of the union bound error probability
of SM-ML detector as {[}\ref{R.-Mesleh,-H.}{]}, {[}\ref{E.-Ba=00015Far,-=0000DC. imperfect csi}{]}

\begin{equation}
P_{b}=\frac{1}{\left(\eta\right)2^{\eta}}\sum_{k=1}^{2^{\eta}}\sum_{l=1}^{2^{\eta}}\delta_{k,l}\mathcal{\mathbb{E}}\left\{ \mathbb{P}\text{r}^{\left(\text{ML}\right)}\left(\mathbf{x}_{k}\rightarrow\tilde{\mathbf{x}}_{l}\right)\right\} ,\label{eq: UB general}
\end{equation}

\noindent where $P_{b}$ is the union bound probability, $\mathbb{P}\text{r}^{\left(\text{ML}\right)}\left(\mathbf{x}_{k}\rightarrow\tilde{\mathbf{x}}_{l}\right)$
stands for the pairwise error probability (PEP) of the proposed SM-ML
decoder, $\delta_{k,l}$ represents the number of bit errors which
corresponds to the instant PEP event, and the spectral efficiency
$\eta$ is given from (\ref{eq: Spectral Efficiency}). Let us consider
that $\Delta^{\text{m-M}}$ is the NoM between the m-M algorithm solution
and the ML solution. Now, the PEP of the m-M algorithm is denoted
by $\mathbb{P}\text{r}^{\left(\text{m-M}\right)}\left(\mathbf{x}_{k}\rightarrow\tilde{\mathbf{x}}_{l}\right)$
and given as {[}\ref{A.-Younis,-S. IEEE Trans}{]}

\begin{equation}
\mathbb{P}\text{r}^{\left(\text{m-M}\right)}\left(\mathbf{x}_{k}\rightarrow\tilde{\mathbf{x}}_{l}\right)=\mathbb{P}\text{r}^{\left(\text{ML}\right)}\left(\mathbf{x}_{k}\rightarrow\tilde{\mathbf{x}}_{l}\right)+\mathbb{P}\text{r}\left(\Delta^{\text{m-M}}\neq0\right).\label{eq: PEP independent}
\end{equation}

\noindent According to (\ref{eq: opt condition}), if the m-M algorithm
detects a minimum ED at the end of fully expanded branch, this means
that no further expansion will happen in the current minimum ED (the
branch length can not be $N_{r}+1$) and the current minimum ED is
a global minimum across all other branches. Therefore, the ML solution
will not be missed (i.e., $\mathbb{P}\text{r}\left(\Delta^{\text{m-M}}\neq0\right)=0$)
and the union bound error probability of the proposed m-M algorithm
is exactly the same as (\ref{eq: UB general}).

\begin{figure}
\begin{centering}
\includegraphics[scale=0.45]{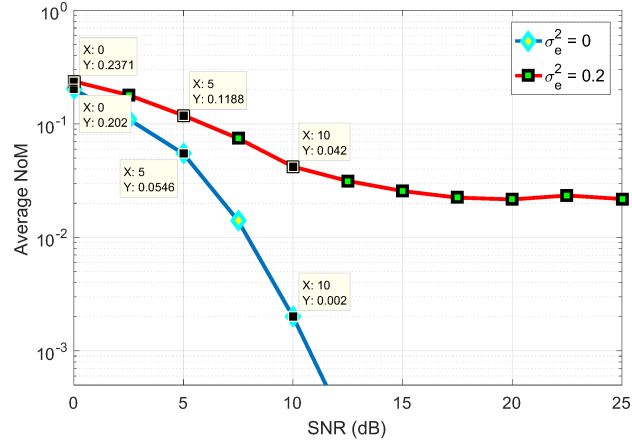}
\par\end{centering}
\caption{{\small{}\label{fig:ML-solution-misses}Average number of NoM of the
m-Mw algorithm for $8\times8$ MIMO-SM and $8$-QAM. }}
\end{figure}

To study the effect of removing the optimality condition in (\ref{eq: opt condition}),
consider an m-M algorithm without this condition, referred to as the
m-Mw algorithm. It should be noted that the m-Mw algorithm is not
a stand-alone algorithm, and it is mentioned here to discuss the optimality
condition in (\ref{eq: opt condition}) for the proposed m-M algorithm.
The m-Mw algorithm stops and declares the solution whenever only one
branch is fully expanded. In such a case, the NoM takes a non-zero
value and $\mathbb{P}\text{r}\left(\Delta^{\text{m-Mw}}\neq0\right)\neq0$.
Fig. \ref{fig:ML-solution-misses} shows the average NoM versus SNR;
$10^{4}$ Rayleigh flat fading channel realizations are run for each
SNR value, for $8\times8$ MIMO-SM using 8-QAM. As we can see, the
NoM reduces as SNR increases and $\sigma_{e}^{2}$ decreases. For
instance, the m-Mw algorithm misses $2020$, $564$ and $20$ ML solution
out of $10^{4}$ runs at SNR of $0$, $5$ and $10$ dB, respectively,
in case of perfect CSIR; for imperfect CSIR with $\sigma_{e}^{2}=0.2$,
the NoM for the m-Mw algorithm is $2371$, $1188$ and $420$ out
of $10^{4}$ runs at SNR of $0$, $5$ and $10$ dB, respectively.

\noindent Hence, the condition in (\ref{eq: opt condition}) ensures
that the minimum ED which comes at the end of a fully expanded branch
is a global minimum across all branches; thus, the ML solution is
achieved. Additionally, omitting the condition in (\ref{eq: opt condition})
leads to a significant BER deterioration when compared with the ML
performance.

\section{\label{sec:Numerical-results}Numerical Results and Discussions}

\begin{figure}[b]
\begin{centering}
\subfloat[8-QAM for $8\times8$ MIMO-SM.]{\includegraphics[scale=0.31]{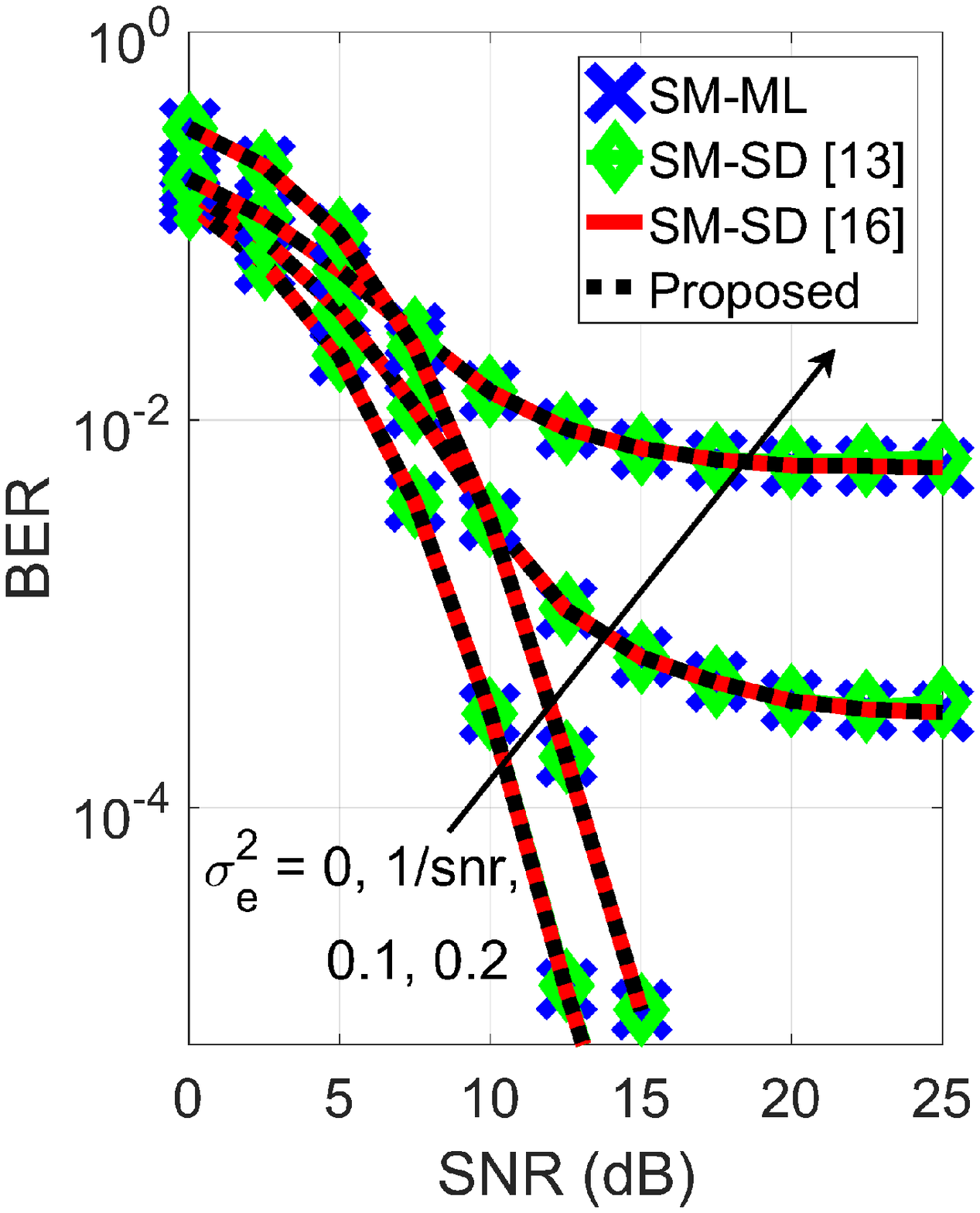}
\centering{}}~\subfloat[16-QAM for $16\times16$ MIMO-SM.]{\includegraphics[scale=0.31]{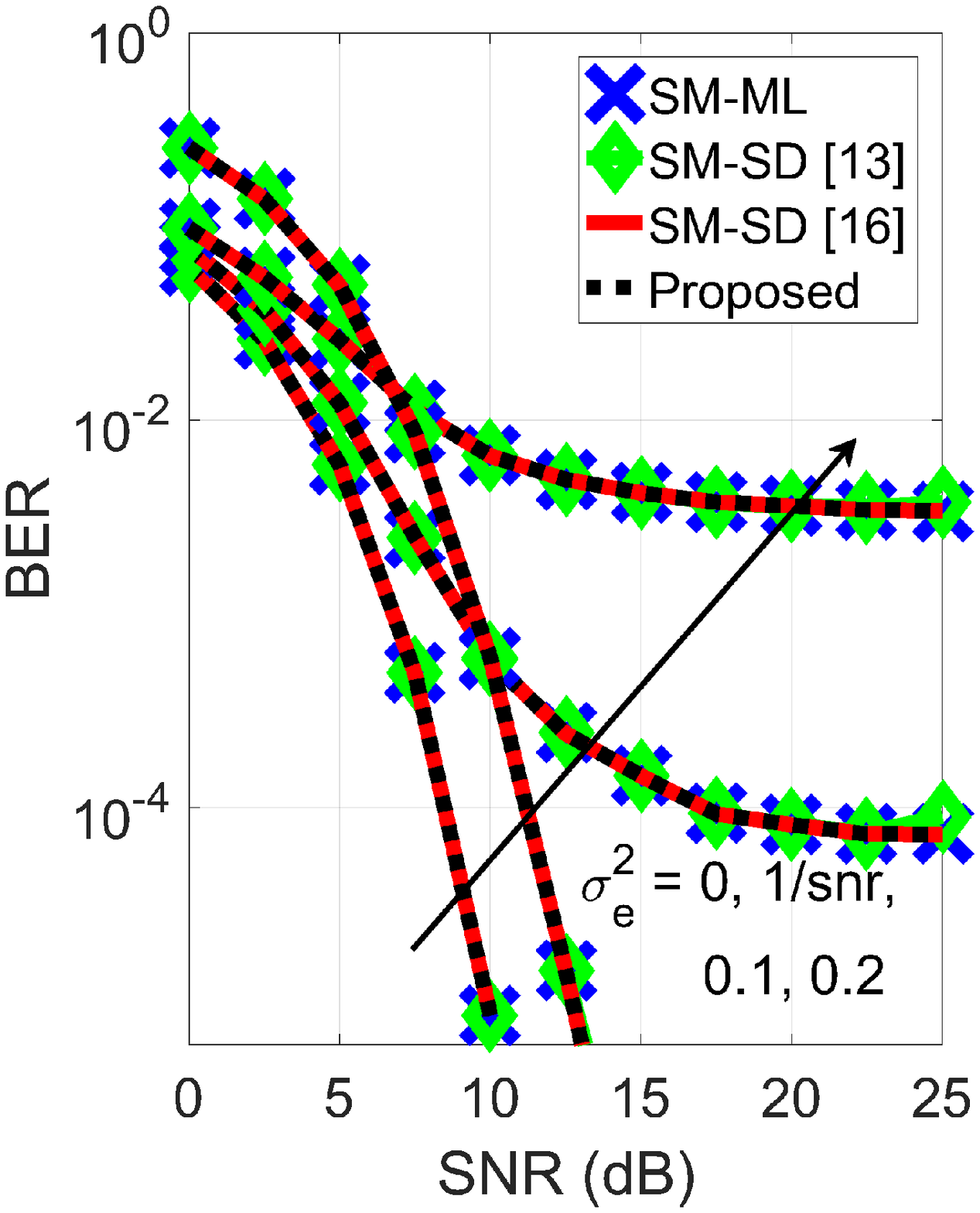}

}
\par\end{centering}
\caption{{\small{}\label{fig:BER-determine}BER comparison of determined MIMO-SM
system for different decoders. }}
\end{figure}

\begin{figure}[b]
\begin{centering}
\subfloat[8-QAM for $6\times8$ MIMO-SM.]{\includegraphics[scale=0.31]{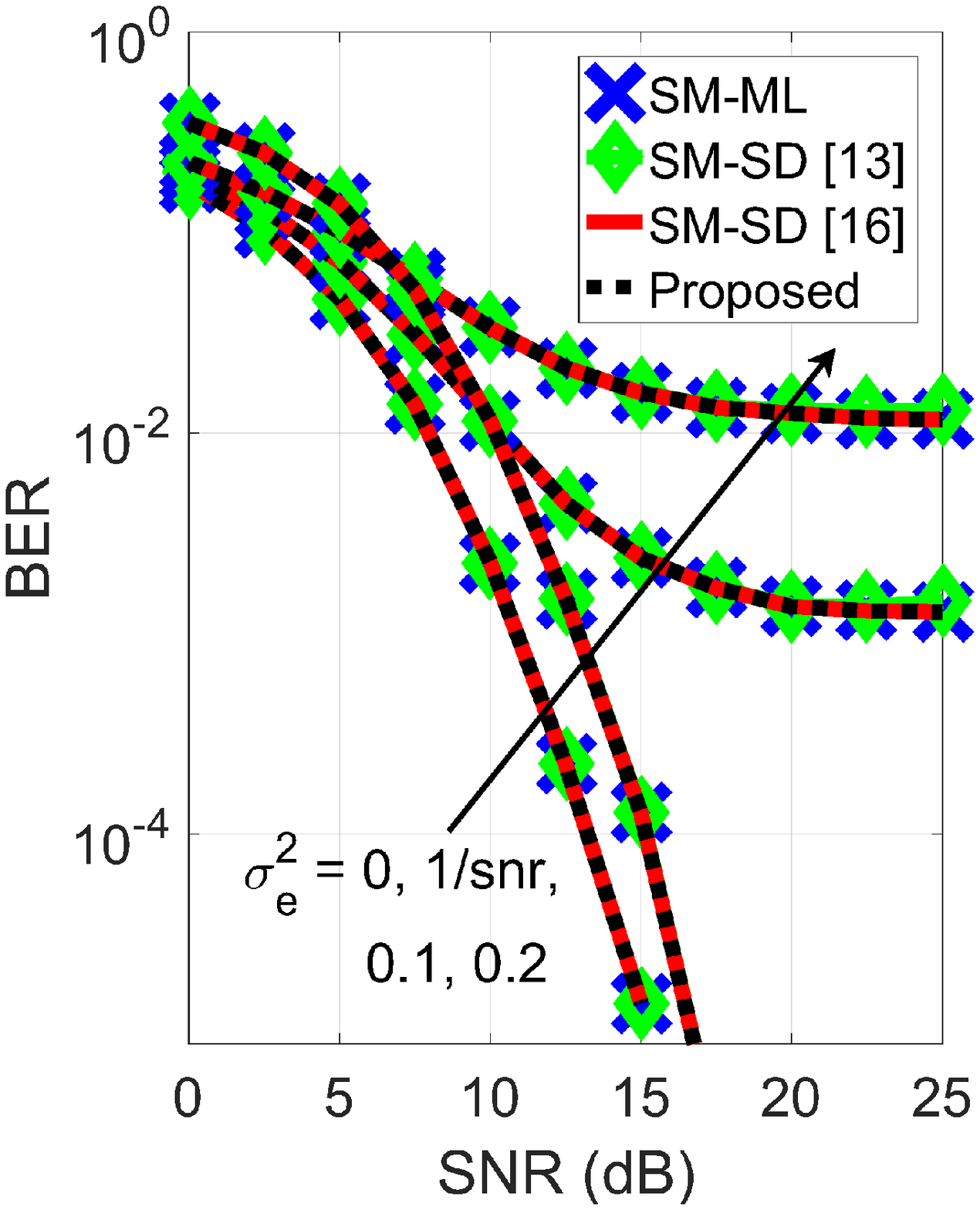}
\centering{}}~\subfloat[16-QAM for $12\times16$ MIMO-SM.]{\includegraphics[scale=0.31]{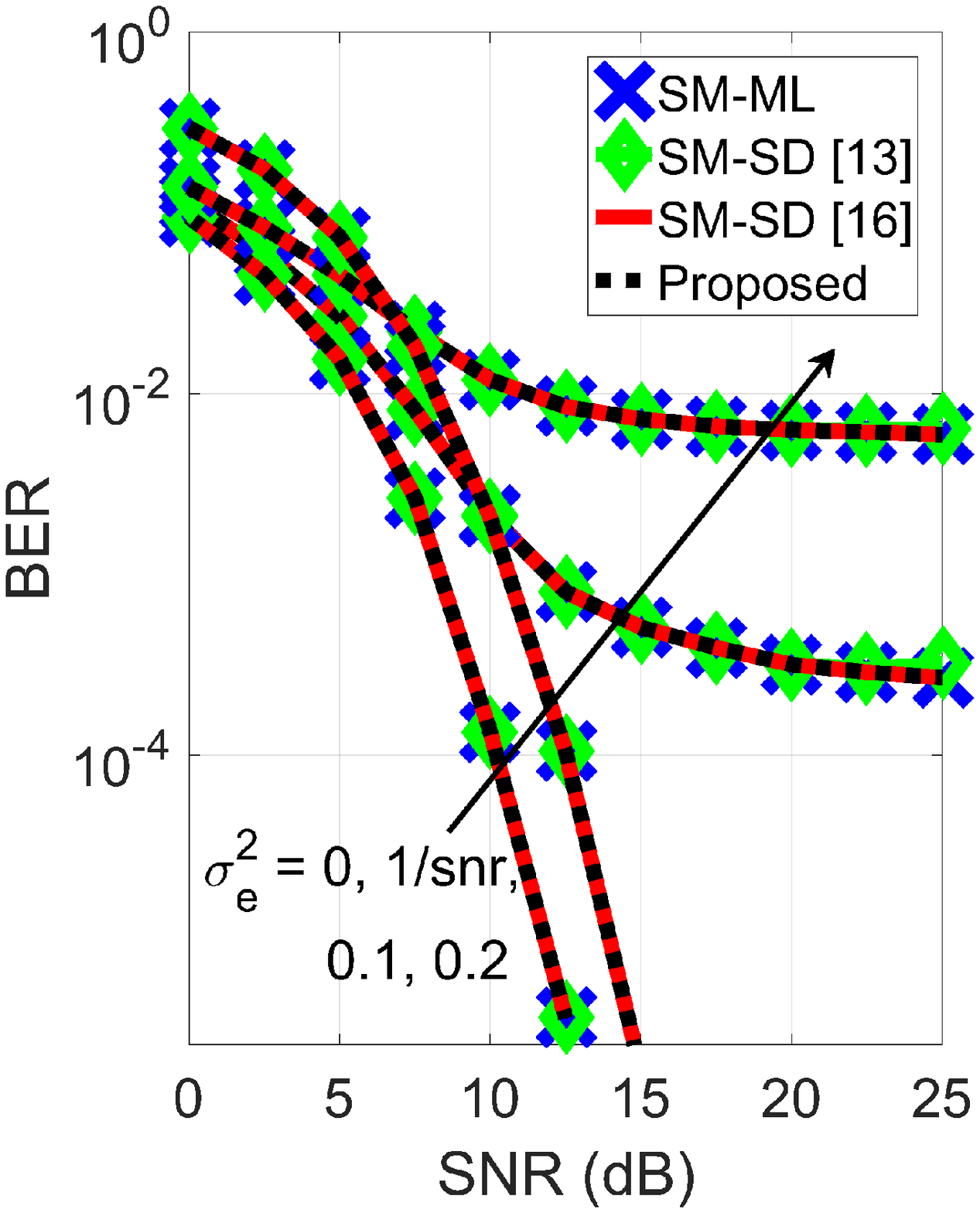}

}
\par\end{centering}
\caption{{\small{}\label{fig:BER-under-determine}BER comparison of under-determined
MIMO-SM system for different decoders. }}
\end{figure}

\begin{figure}
\begin{centering}
\subfloat[8-QAM for $10\times8$ MIMO-SM.]{\includegraphics[scale=0.31]{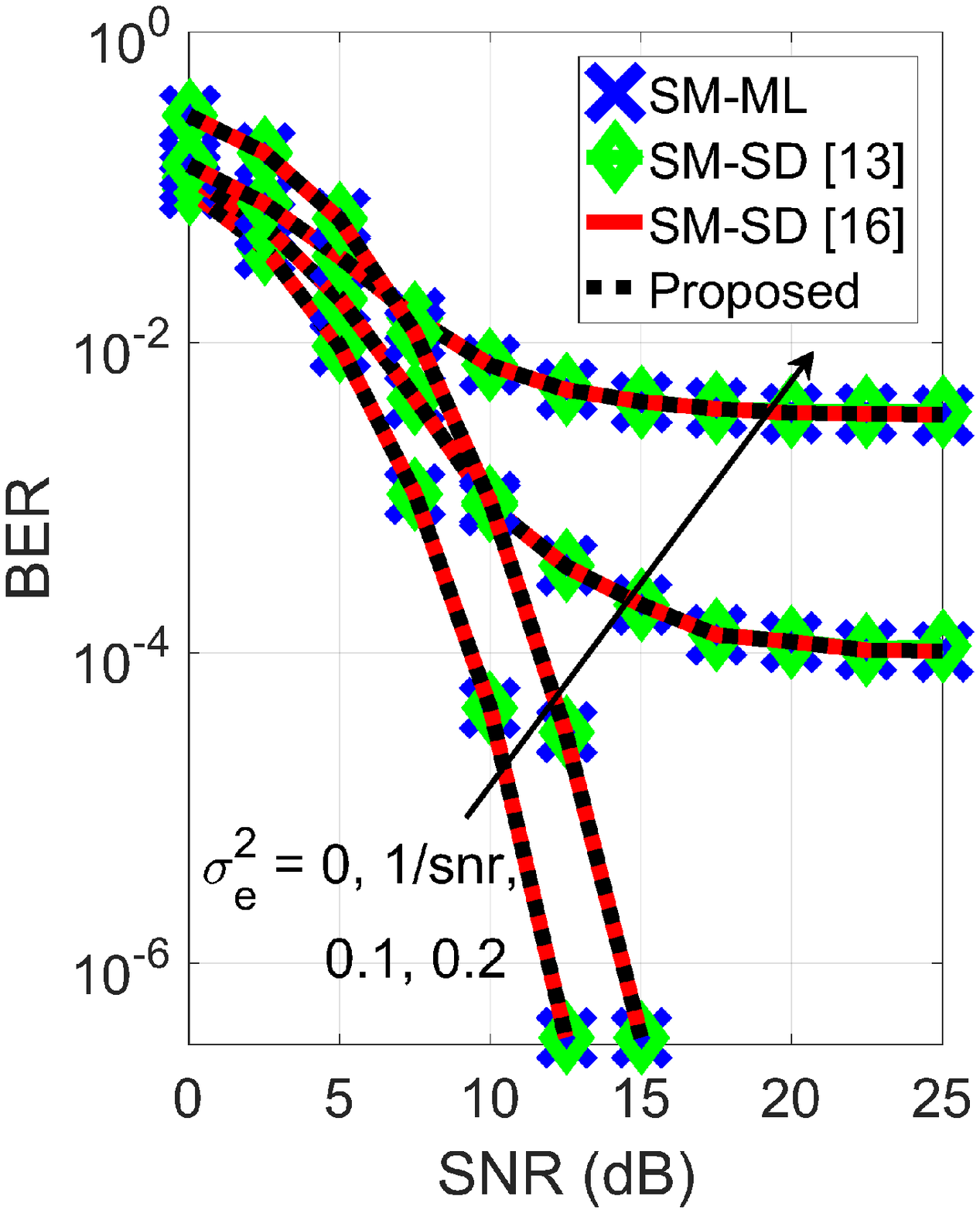}
\centering{}}~\subfloat[16-QAM for $20\times16$ MIMO-SM.]{\includegraphics[scale=0.31]{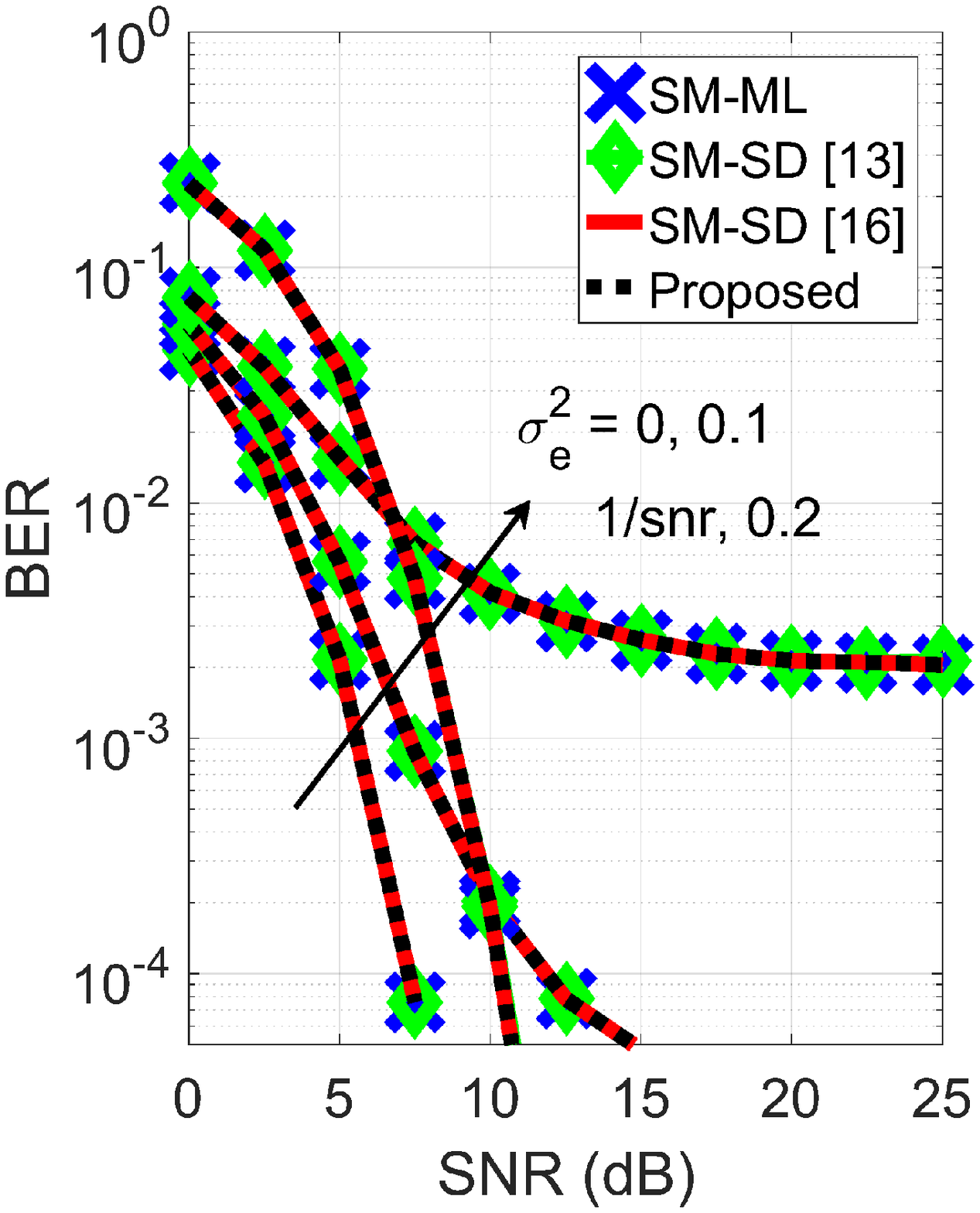}

}
\par\end{centering}
\caption{{\small{}\label{fig:BER-over-determine}BER comparison of over-determined
MIMO-SM system for different decoders. }}
\end{figure}

In this section, we evaluate the behavior of the proposed m-M algorithm
in terms of BER and decoding complexity. In addition, comparisons
between the m-M algorithm and SM-SD algorithms in the literature are
presented. Since the m-M algorithm provides the optimal BER performance,
we consider the SM-SD algorithms (such as given in {[}\ref{A.-Younis,-S. IEEE Trans}{]}
and {[}\ref{I.-Al-Nahhal,-O. QSM}{]}) in comparisons. Three scenarios
are considered: a) perfect CSIR ($\sigma_{e}^{2}=0$), b) imperfect
CSIR with fixed $\sigma_{e}^{2}$ ($\sigma_{e}^{2}=0.1$ and $0.2$),
and c) imperfect CSIR with variable $\sigma_{e}^{2}$ ($\sigma_{e}^{2}=1/\text{snr}$).
Two spectral efficiency values are considered: $\eta=6$ bpcu using
8-QAM for $N_{r}\times8$ MIMO-SM system, and $\eta=8$ bpcu using
16-QAM for $N_{r}\times16$ MIMO-SM system. The value of $N_{r}$
for both cases describes the type of the system. In the case of determined
MIMO-SM system, $N_{r}=N_{t}$ (i.e., $N_{r}=8$ and $16$ for $\eta=6$
and 8, respectively). For under-determined MIMO-SM system, $N_{r}<N_{t}$
(e.g., $N_{r}=6$ and $12$ for $\eta=6$ and $8$, respectively).
Finally, we have an over-determined MIMO-SM system if $N_{r}>N_{t}$
(e.g., $N_{r}=10$ and $20$ for $\eta=6$ and $8$, respectively).
Monte Carlo simulations are used to obtain the presented results for
all scenarios by running at least $5\times10^{5}$ Rayleigh flat fading
channel realizations.

\subsection{BER Comparison}

In this subsection, the BER performance of the SM-ML, SM-SD {[}\ref{A.-Younis,-S. IEEE Trans}{]},
SM-SD {[}\ref{I.-Al-Nahhal,-O. QSM}{]}, and proposed m-M algorithms
are compared with respect to SNR. Figs. \ref{fig:BER-determine},
\ref{fig:BER-under-determine} and \ref{fig:BER-over-determine} show
the BER performance of different SM decoders for determined, under-determined,
and over-determined MIMO-SM systems, respectively. The left sub-plots
in all three figures present $\eta=6$ bpcu, while the right ones
show $\eta=8$ bpcu. As observed from these figures, the two SM-SD
algorithms in {[}\ref{A.-Younis,-S. IEEE Trans}{]} and {[}\ref{I.-Al-Nahhal,-O. QSM}{]},
as well as the proposed m-M algorithm provide the same SM-ML BER for
all values of $\sigma_{e}^{2}$ (i.e., 0, 0.1, 0.2, and 1/snr). As
expected, the best BER is obtained when $\sigma_{e}^{2}=0$, while
the BER degrades for increasing values of $\sigma_{e}^{2}$. Unlike
the BER obtained from having $\sigma_{e}^{2}=1/\text{snr}$, an error
floor occurs in the case of $\sigma_{e}^{2}=0.1$ and 0.2 even in
high SNR due to the fixed values of $\sigma_{e}^{2}$. The error floor
is mitigated as $N_{r}$ increases. For instance, the error floor
of the $\sigma_{e}^{2}=0.1$ curve in Fig. \ref{fig:BER-under-determine}(b)
can not be reduced to $5\times10^{-4}$ when $N_{r}=12$; when $N_{r}=16$
in Fig. \ref{fig:BER-determine}(b) for the $\sigma_{e}^{2}=0.1$
curve, the error floor occurs at $10^{-4}$; however, it further reduces
to $10^{-5}$ when $N_{r}=20$ in Fig. \ref{fig:BER-over-determine}(b)
for the $\sigma_{e}^{2}=0.1$ curve.

It can be seen from these figures that there is no preference in BER
between the proposed m-M algorithm and the other SM-SD algorithms
in {[}\ref{A.-Younis,-S. IEEE Trans}{]} and {[}\ref{I.-Al-Nahhal,-O. QSM}{]}.
For all presented scenarios, the low-complexity algorithms (the m-M
algorithm, and SM-SD algorithms in {[}\ref{A.-Younis,-S. IEEE Trans}{]}
and {[}\ref{I.-Al-Nahhal,-O. QSM}{]}) provide the same BER as the
SM-ML detection. It is worth noting that in practice, the channel
estimation accuracy improves as the SNR increases (i.e., $\sigma_{e}^{2}=1/\mathrm{snr}$),
and the ML BER performance can be still reasonable, as seen from Figs.
\ref{fig:BER-determine}, \ref{fig:BER-under-determine} and \ref{fig:BER-over-determine}.

\begin{figure}
\begin{centering}
\subfloat[{\small{}8-QAM for $8\times8$ MIMO-SM.}]{\begin{centering}
\includegraphics[scale=0.36]{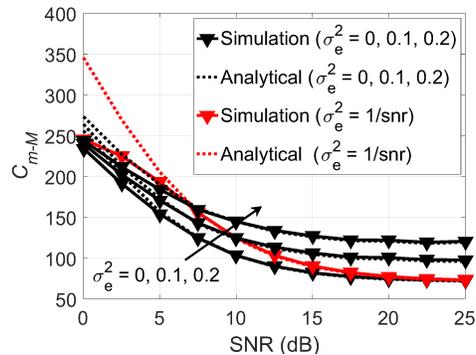}
\par\end{centering}
\centering{}}
\par\end{centering}
\begin{centering}
\subfloat[{\small{}16-QAM for $16\times16$ MIMO-SM.}]{\begin{centering}
\includegraphics[scale=0.36]{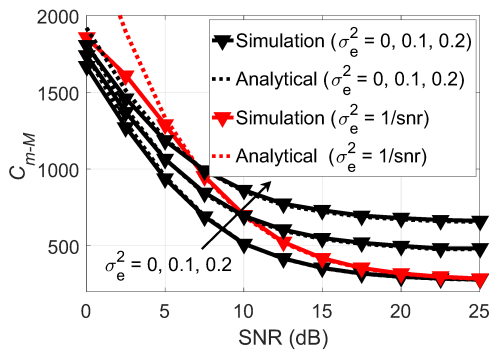}
\par\end{centering}
}
\par\end{centering}
\caption{{\small{}\label{fig:Complexity-of-determined}Complexity of determined
MIMO-SM system for the proposed m-M algorithm.}}
\end{figure}

\begin{figure}
\begin{centering}
\subfloat[{\small{}8-QAM for $6\times8$ MIMO-SM.}]{\begin{centering}
\includegraphics[scale=0.36]{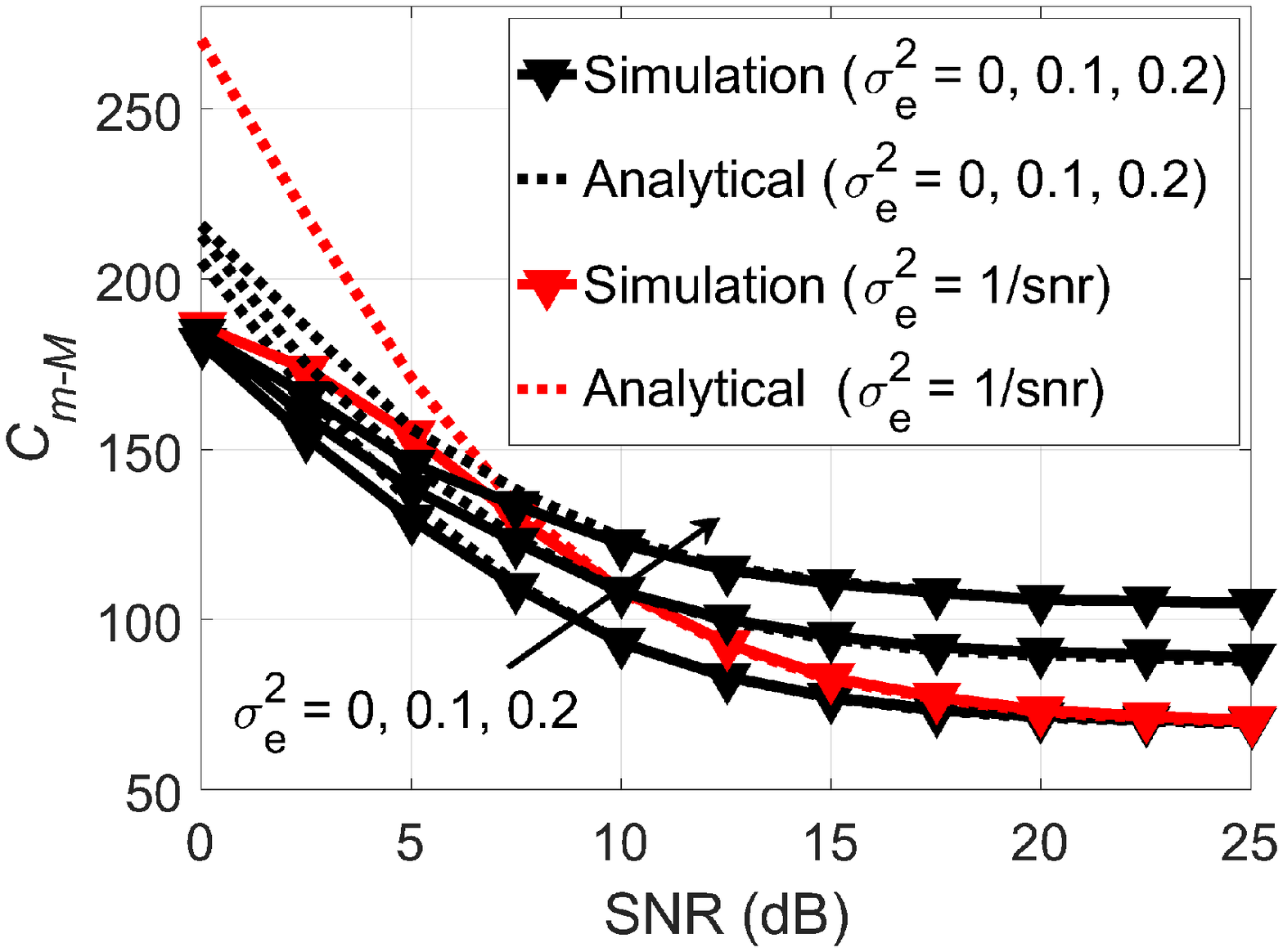}
\par\end{centering}
\centering{}}
\par\end{centering}
\begin{centering}
\subfloat[{\small{}16-QAM for $12\times16$ MIMO-SM.}]{\begin{centering}
\includegraphics[scale=0.36]{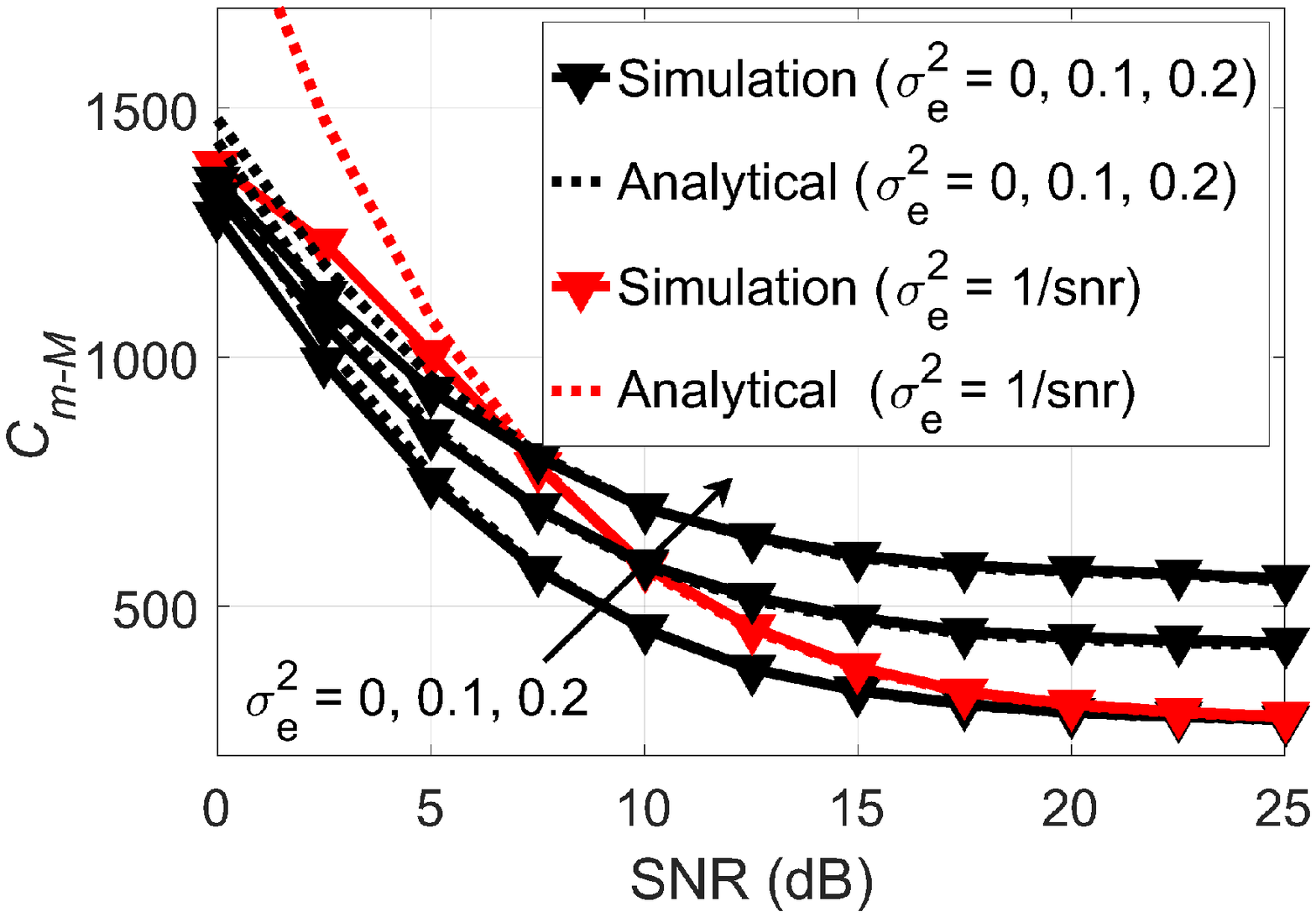}
\par\end{centering}
}
\par\end{centering}
\caption{{\small{}\label{fig:Complexity-of-under}Complexity of under-determined
MIMO-SM system for the proposed m-M algorithm.}}
\end{figure}

\begin{figure}
\begin{centering}
\subfloat[{\small{}8-QAM for $10\times8$ MIMO-SM.}]{\begin{centering}
\includegraphics[scale=0.36]{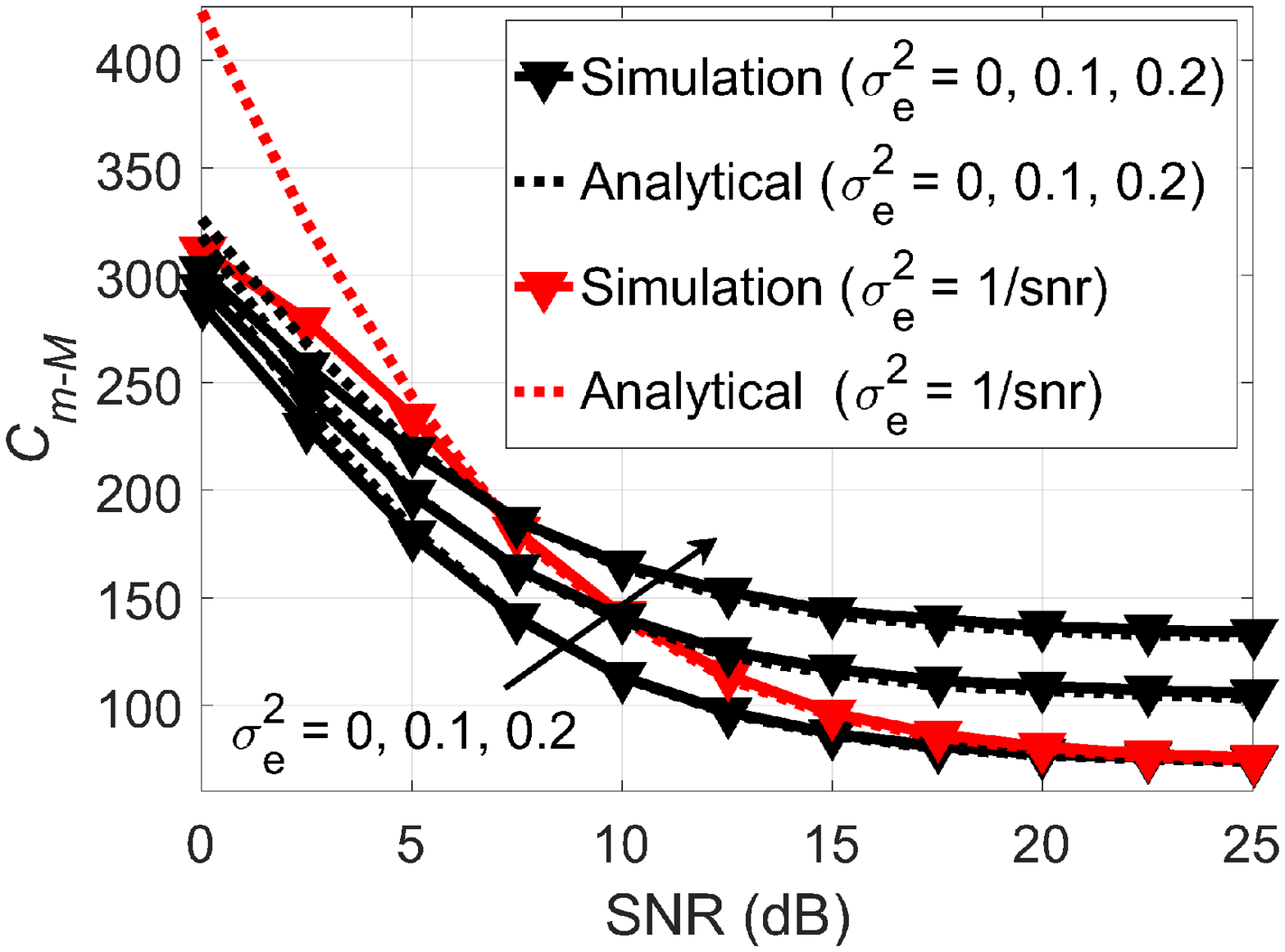}
\par\end{centering}
\centering{}}
\par\end{centering}
\begin{centering}
\subfloat[{\small{}16-QAM for $20\times16$ MIMO-SM.}]{\begin{centering}
\includegraphics[scale=0.36]{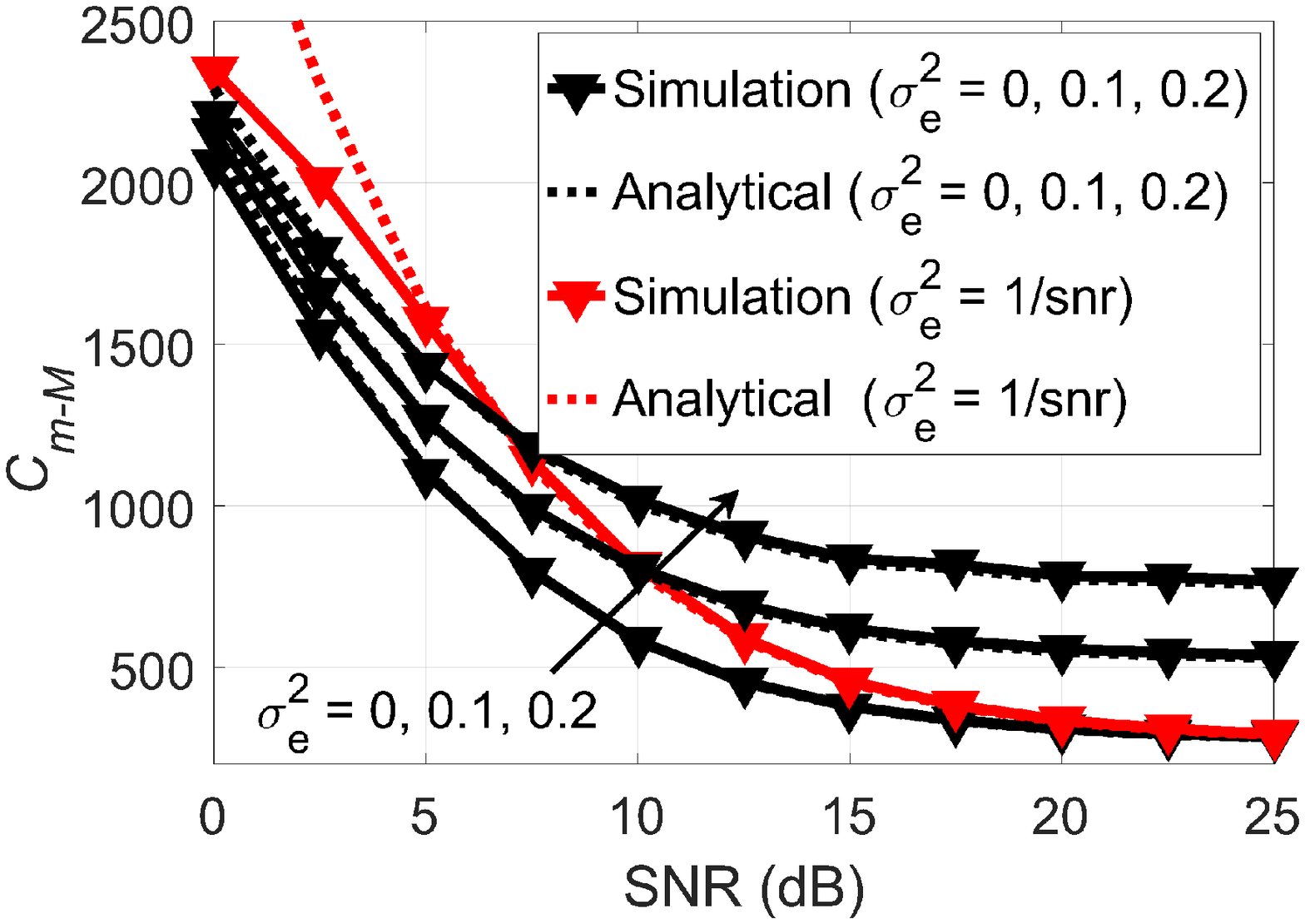}
\par\end{centering}
}
\par\end{centering}
\caption{{\small{}\label{fig:Complexity-of-over}Complexity of over-determined
MIMO-SM system for the proposed m-M algorithm.}}
\end{figure}

\subsection{Analytical Complexity Assessment}

In this subsection, we evaluate the accuracy of the analytical expressions
for the expected decoding complexity of the m-M algorithm given in
(\ref{eq:C_m_M PCSI}) and (\ref{eq:C_m_M imperfect CSI}). As mentioned
before, the number of visited nodes (VNs) is used as a measure for
the decoding complexity of all algorithms in this paper. Figs. \ref{fig:Complexity-of-determined},
\ref{fig:Complexity-of-under} and \ref{fig:Complexity-of-over} present
the comparison results between the analytical expressions and computer
simulation results of the determined, under-determined and over-determined
MIMO-SM systems, respectively, for $\eta=6$ and $8$ bpcu. In all
figures, the analytical expression for $\sigma_{e}^{2}=0$ is given
from (\ref{eq:C_m_M PCSI}), while the analytical expression for $\sigma_{e}^{2}=0.1$,
$0.2$ and $\left(1/\text{snr}\right)$ is given from (\ref{eq:C_m_M imperfect CSI}).
From these figures, we observe that the analytical expressions in
(\ref{eq:C_m_M PCSI}) and (\ref{eq:C_m_M imperfect CSI}) match the
computer simulation results after SNR values of 5 dB, while some mismatches
occur at low SNR values.

It should be noted that the mismatch between the analytical expressions
and simulation results at low SNR values comes from the assumptions
of $\hat{\mathbf{x}}_{\text{m-M}}\rightarrow\mathbf{x}_{t}$ in (\ref{eq:R_m_M assumption}).
At low SNR, the ML solution (the same as $\hat{\mathbf{x}}_{\text{m-M}}$)
misses the true solution, $\mathbf{x}_{t}$, which means that $\left\Vert \mathbf{y}-\mathbf{x}_{t}\right\Vert _{F}^{2}>\left\Vert \mathbf{y}-\hat{\mathbf{x}}_{\text{m-M}}\right\Vert _{F}^{2}$.
In other words, the threshold $R_{\text{m-M}}$ in (\ref{eq:R_m_M assumption})
used for the analytical expressions will be greater than the actual
threshold in (\ref{eq: R_m_M}), which leads to the count of more
nodes than the reality. By increasing the SNR, the ML solution most
probably estimates the true solution; the assumption of $\hat{\mathbf{x}}_{\text{m-M}}\rightarrow\mathbf{x}_{t}$
becomes more reliable. In the case of $\sigma_{e}^{2}=1/\text{snr}$,
$\sigma_{e}^{2}$ becomes very high at low SNR values (e.g., $\sigma_{e}^{2}=1$
at zero SNR) which dramatically affects the accuracy of (\ref{eq:C_m_M imperfect CSI}).

As it can be seen from these figures, the derived analytical expressions
in (\ref{eq:C_m_M PCSI}) and (\ref{eq:C_m_M imperfect CSI}) accurately
describe the decoding complexity of the proposed m-M algorithm in
both perfect and imperfect CSIR especially at high SNR values for
determined, under-determined, and over-determined MIMO-SM systems.

\subsection{Complexity Comparison}

In this subsection, we compare the complexity of the proposed m-M
algorithm with the optimal BER performance SM-SD algorithms ({[}\ref{A.-Younis,-S. IEEE Trans}{]}
and {[}\ref{I.-Al-Nahhal,-O. QSM}{]}). It should be noted that the
threshold of the SM-SD algorithm in {[}\ref{A.-Younis,-S. IEEE Trans}{]}
is optimized to provide the optimal BER. The comparison goal is to
determine the decoding complexity reduction ratio between the desired
and SM-ML algorithms, which is given as

\begin{equation}
C_{\text{R}}=\frac{MN_{t}N_{r}-C_{\Lambda}}{MN_{t}N_{r}}=1-\frac{C_{\Lambda}}{MN_{t}N_{r}},\label{eq: Comp Reduction}
\end{equation}

\noindent where $C_{\text{R}}$ denotes the complexity reduction ratio,
$MN_{t}N_{r}$ is the decoding complexity of the ML detector, and
$C_{\Lambda}$ denotes the decoding complexity of the target algorithm
with $\Lambda\in\left\{ \text{m-M},\,\,\text{SM-SD}\,\text{[13]},\,\,\text{SM-SD}\,\text{[16]}\right\} $.
The minimum number of nodes that can be visited by any algorithm is
a one fully expanded branch (i.e., $N_{r}$ nodes) in addition to
the nodes of the first row in the tree-search (i.e., $MN_{t}-1$ nodes).
Thus, we can define the maximum reduction in the decoding complexity
ratio that can be achieved by any algorithm, $C_{\text{R}}^{\text{max}}$,
as 

\begin{equation}
C_{\text{R}}^{\text{max}}=1-\frac{N_{r}+MN_{t}-1}{MN_{t}N_{r}}.\label{eq: Comp Reduction max}
\end{equation}

\begin{figure}
\begin{centering}
\subfloat[$\sigma_{e}^{2}=0$.]{\includegraphics[scale=0.31]{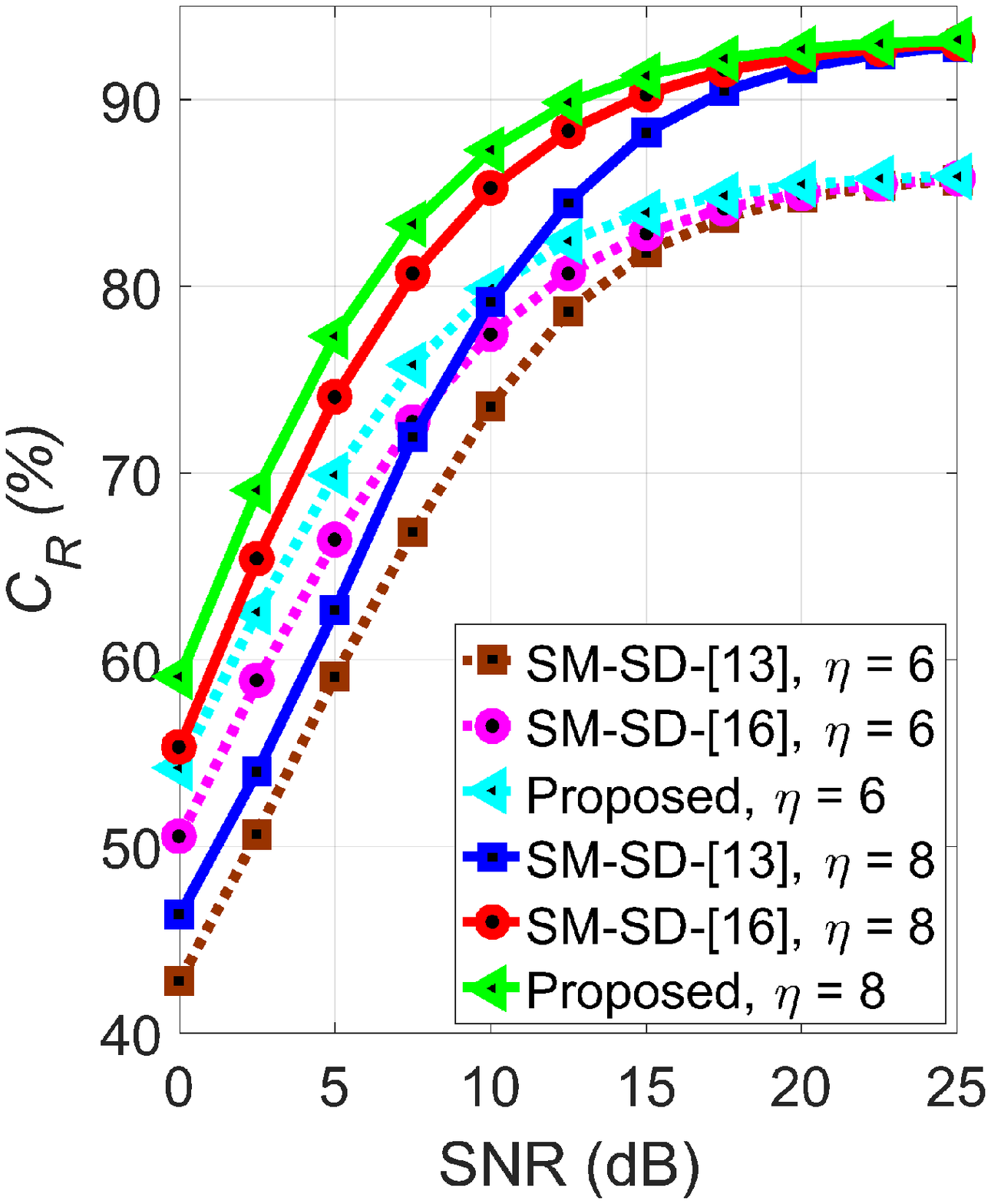}
\centering{}}~\subfloat[$\sigma_{e}^{2}=1/\text{snr}$.]{\includegraphics[scale=0.31]{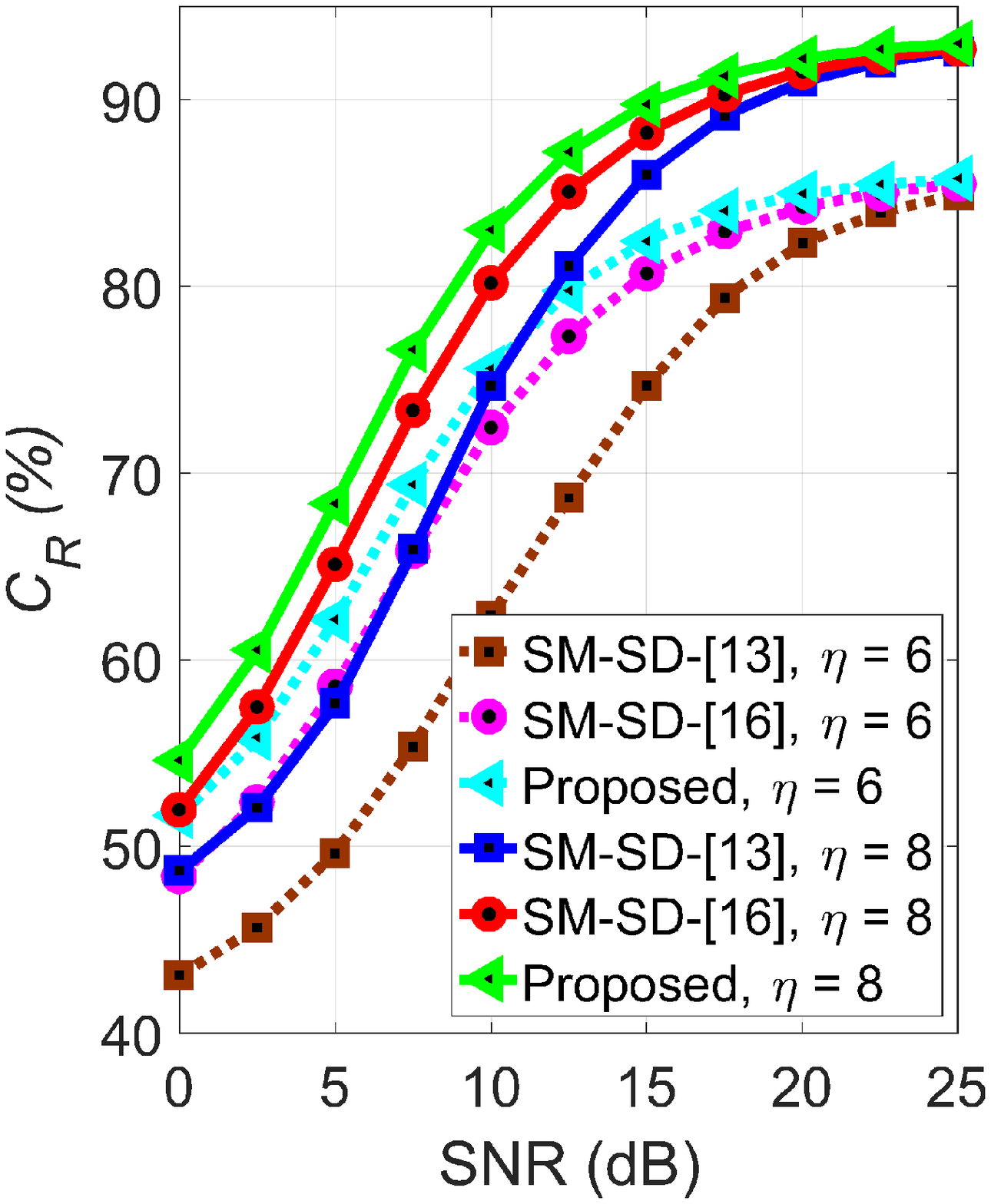}

}
\par\end{centering}
\begin{centering}
\subfloat[$\sigma_{e}^{2}=0.1$.]{\includegraphics[scale=0.31]{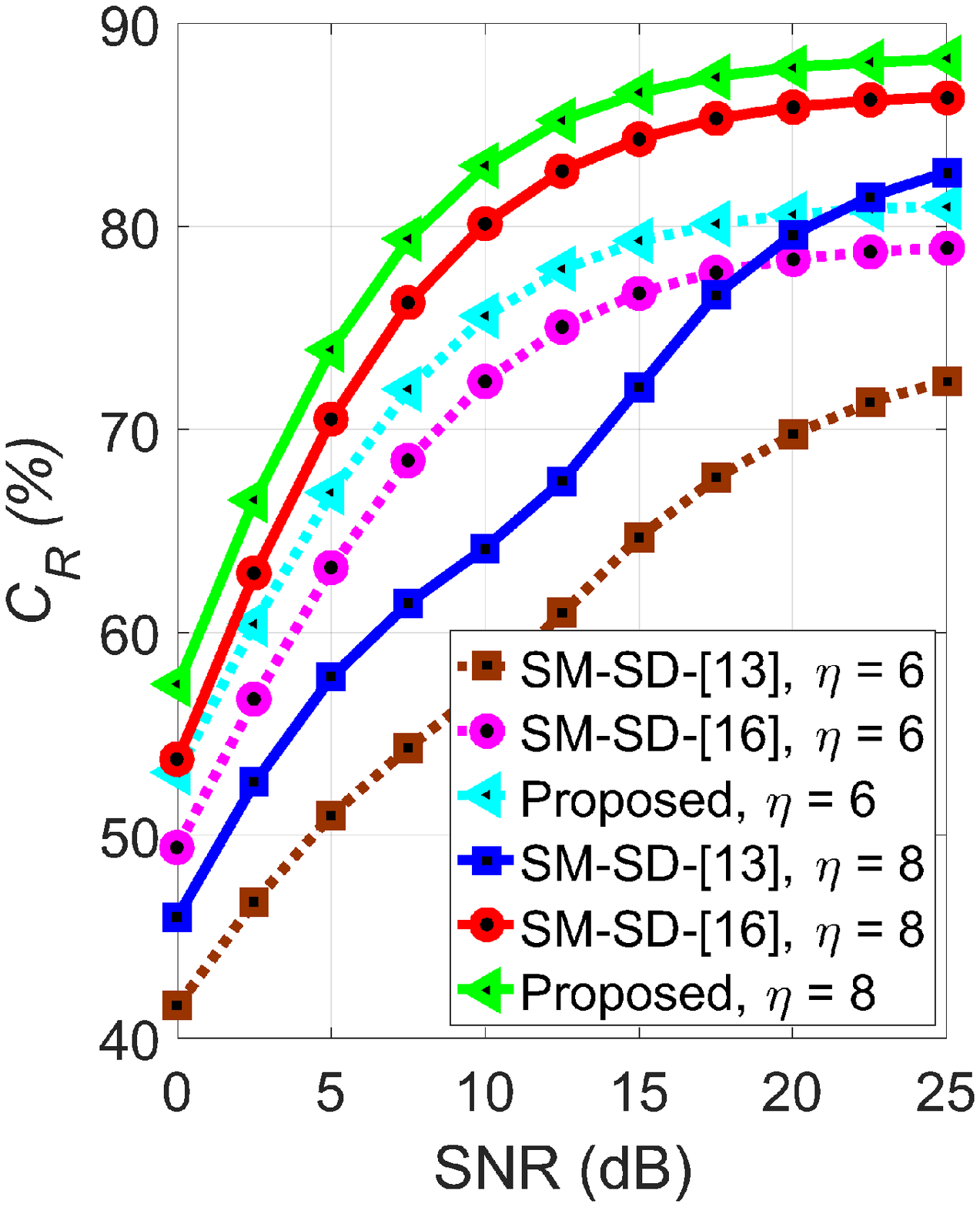}
\centering{}}~\subfloat[$\sigma_{e}^{2}=0.2$.]{\includegraphics[scale=0.31]{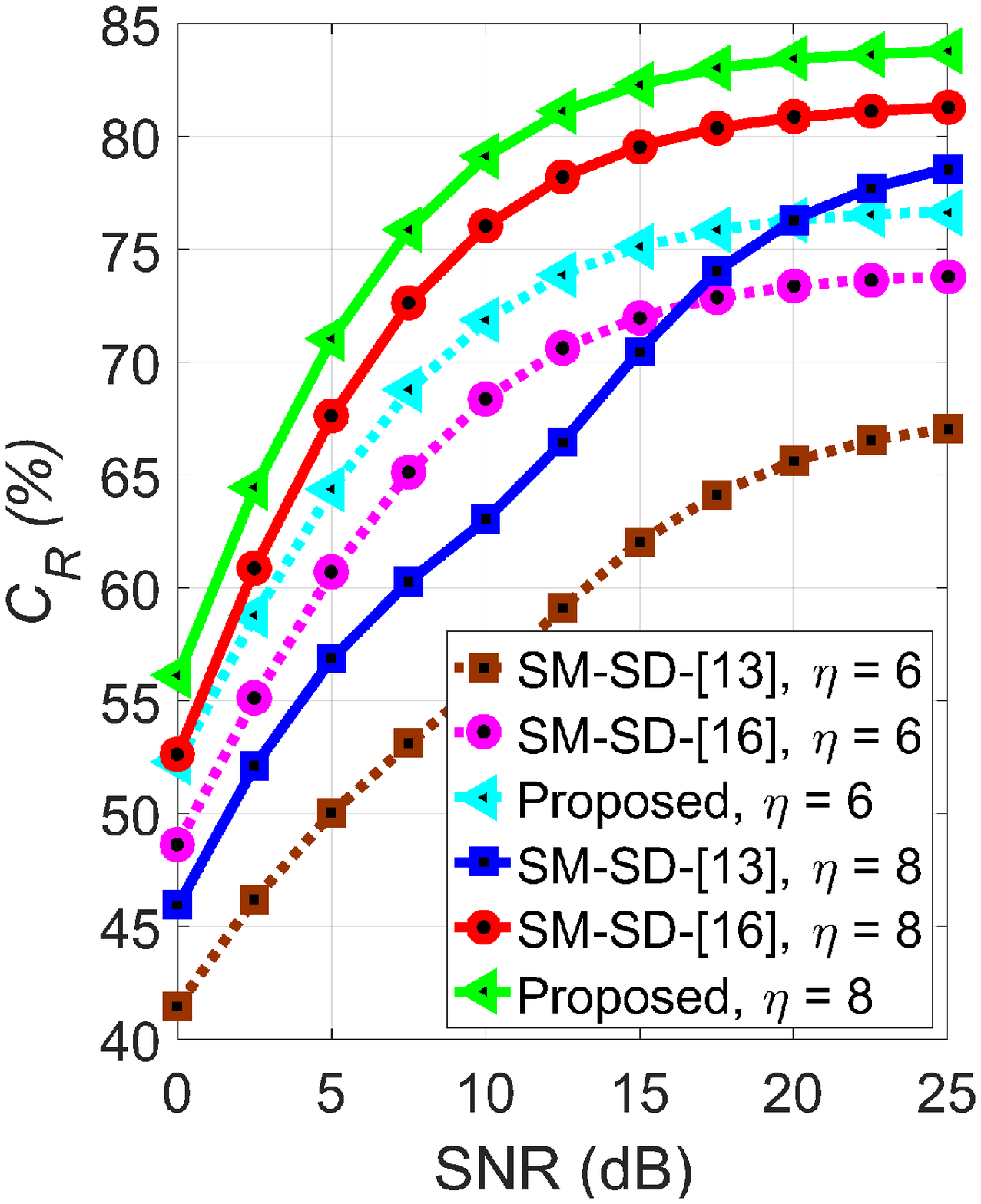}

}
\par\end{centering}
\caption{{\small{}\label{fig:Complexity-reduction-determined}Complexity reduction
comparison of determined MIMO-SM system for different decoders.}}
\end{figure}

\begin{figure}
\begin{centering}
\subfloat[$\sigma_{e}^{2}=0$.]{\includegraphics[scale=0.31]{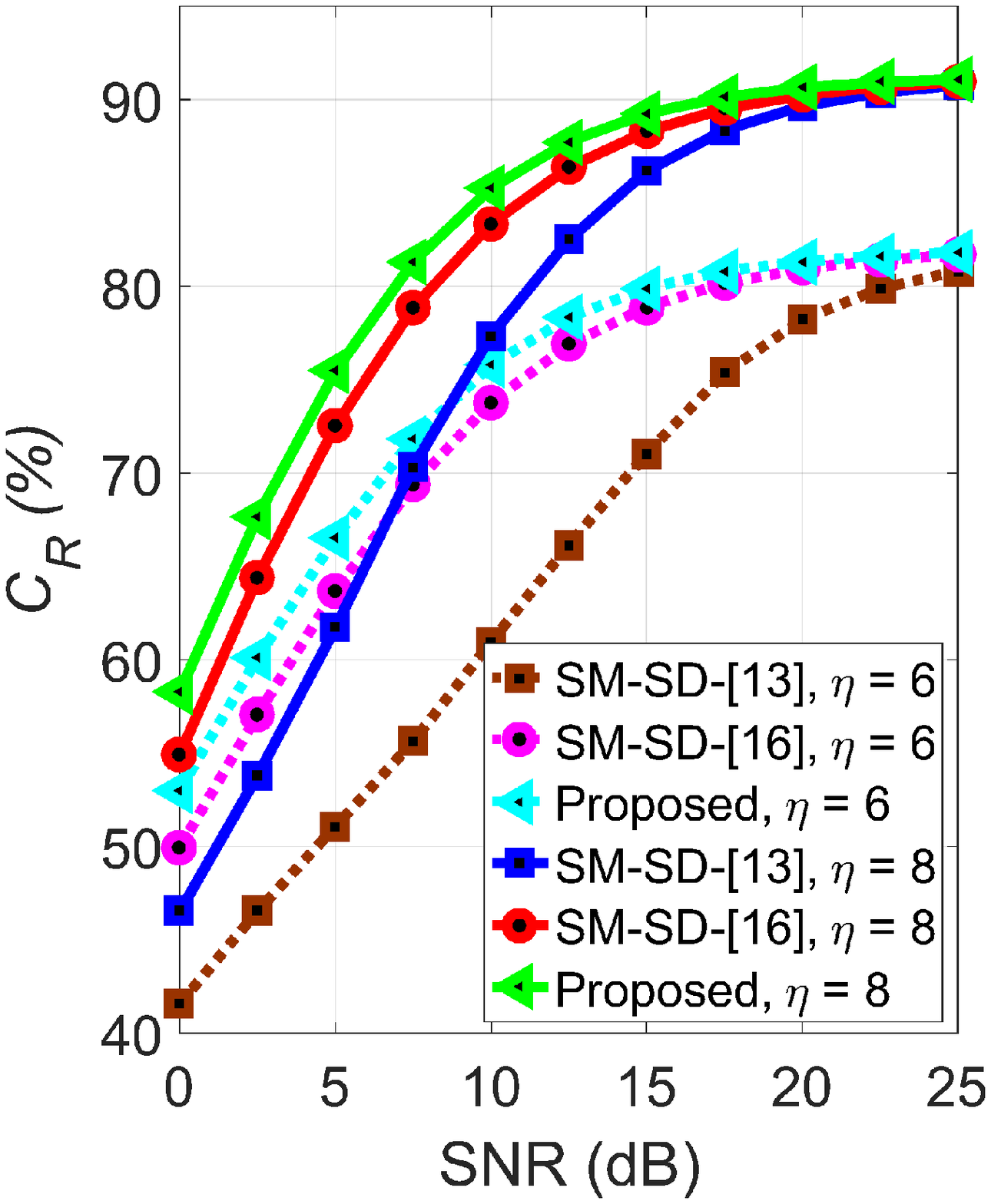}
\centering{}}~\subfloat[$\sigma_{e}^{2}=1/\text{snr}$.]{\includegraphics[scale=0.31]{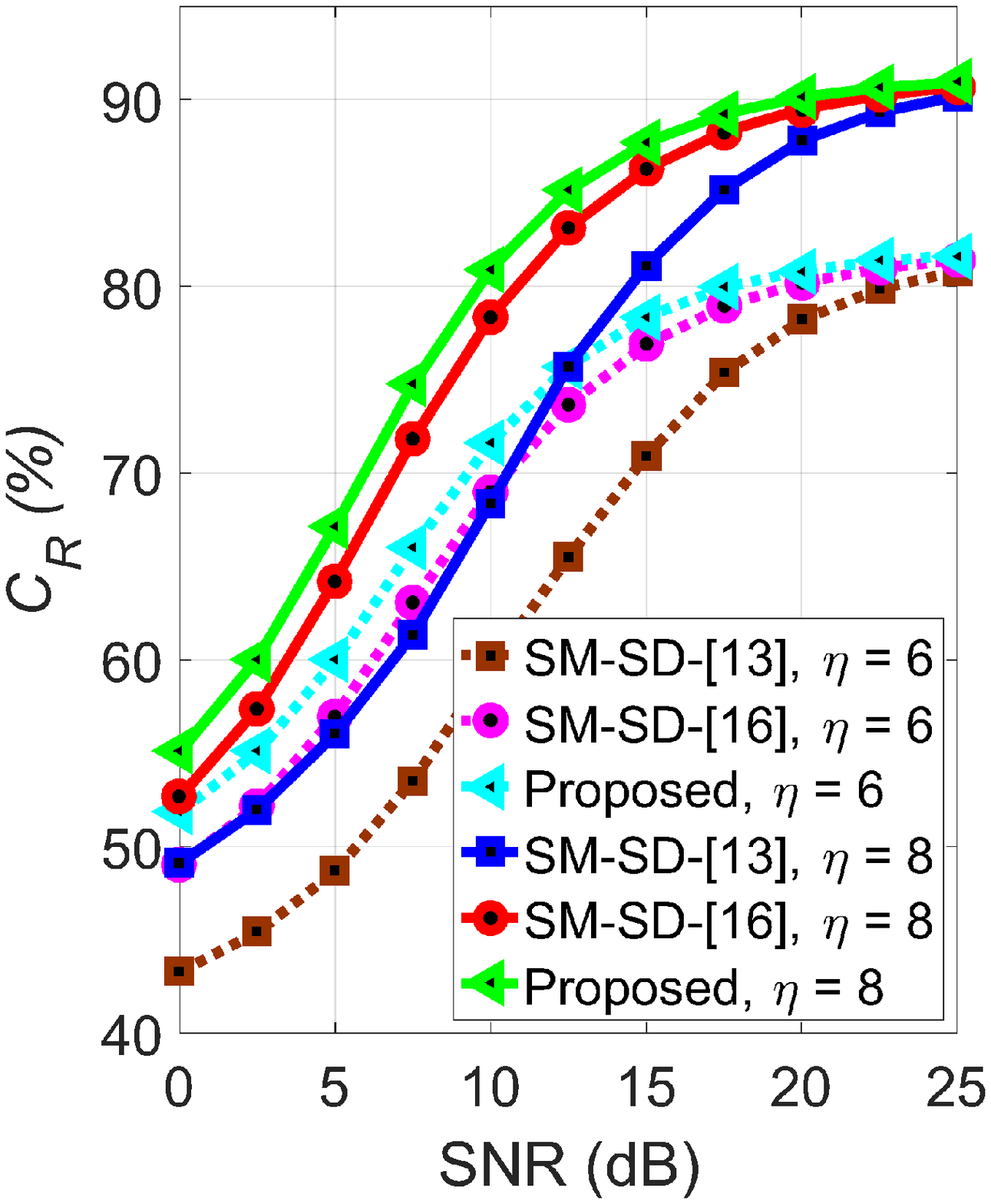}

}
\par\end{centering}
\begin{centering}
\subfloat[$\sigma_{e}^{2}=0.1$.]{\includegraphics[scale=0.31]{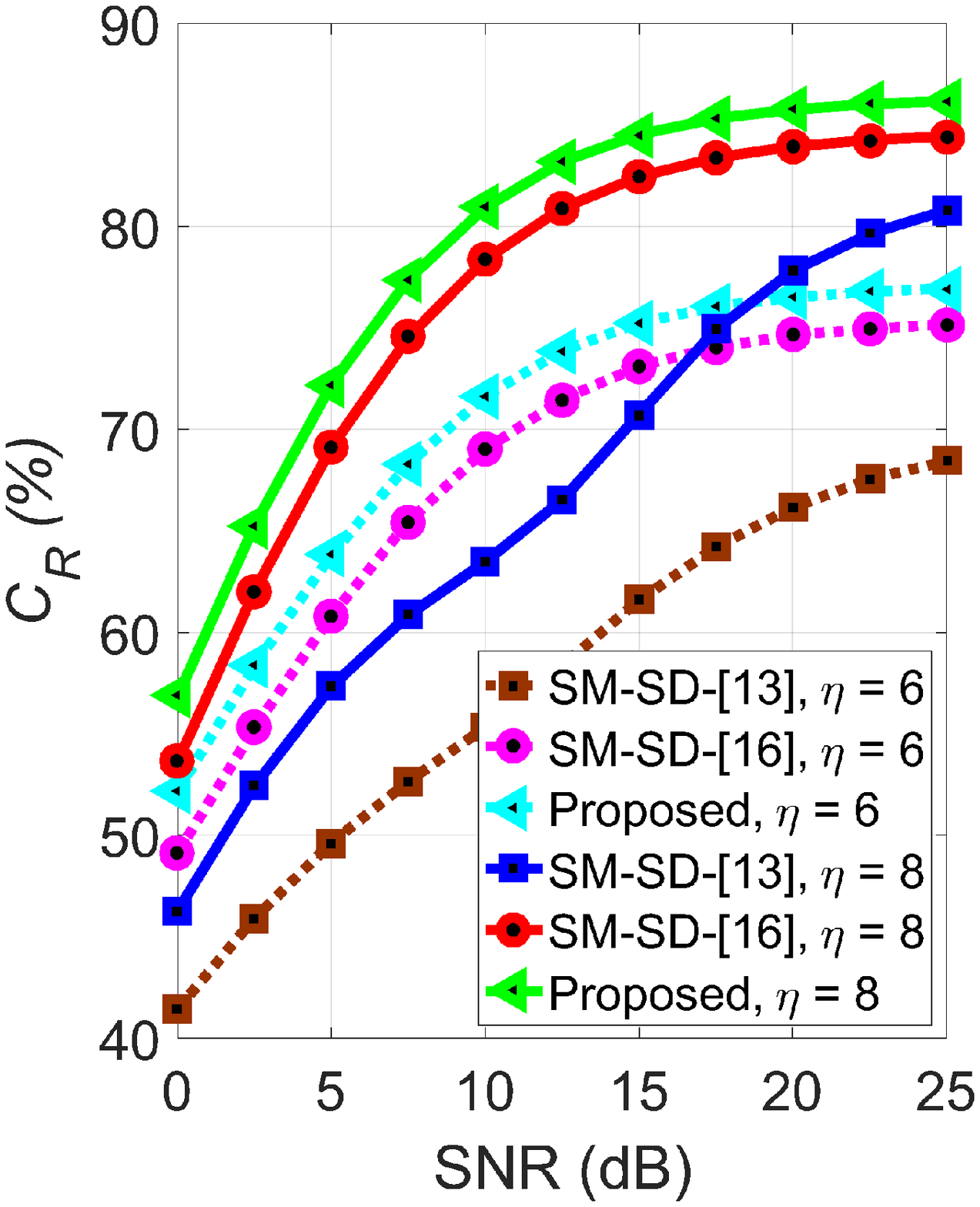}
\centering{}}~\subfloat[$\sigma_{e}^{2}=0.2$.]{\includegraphics[scale=0.31]{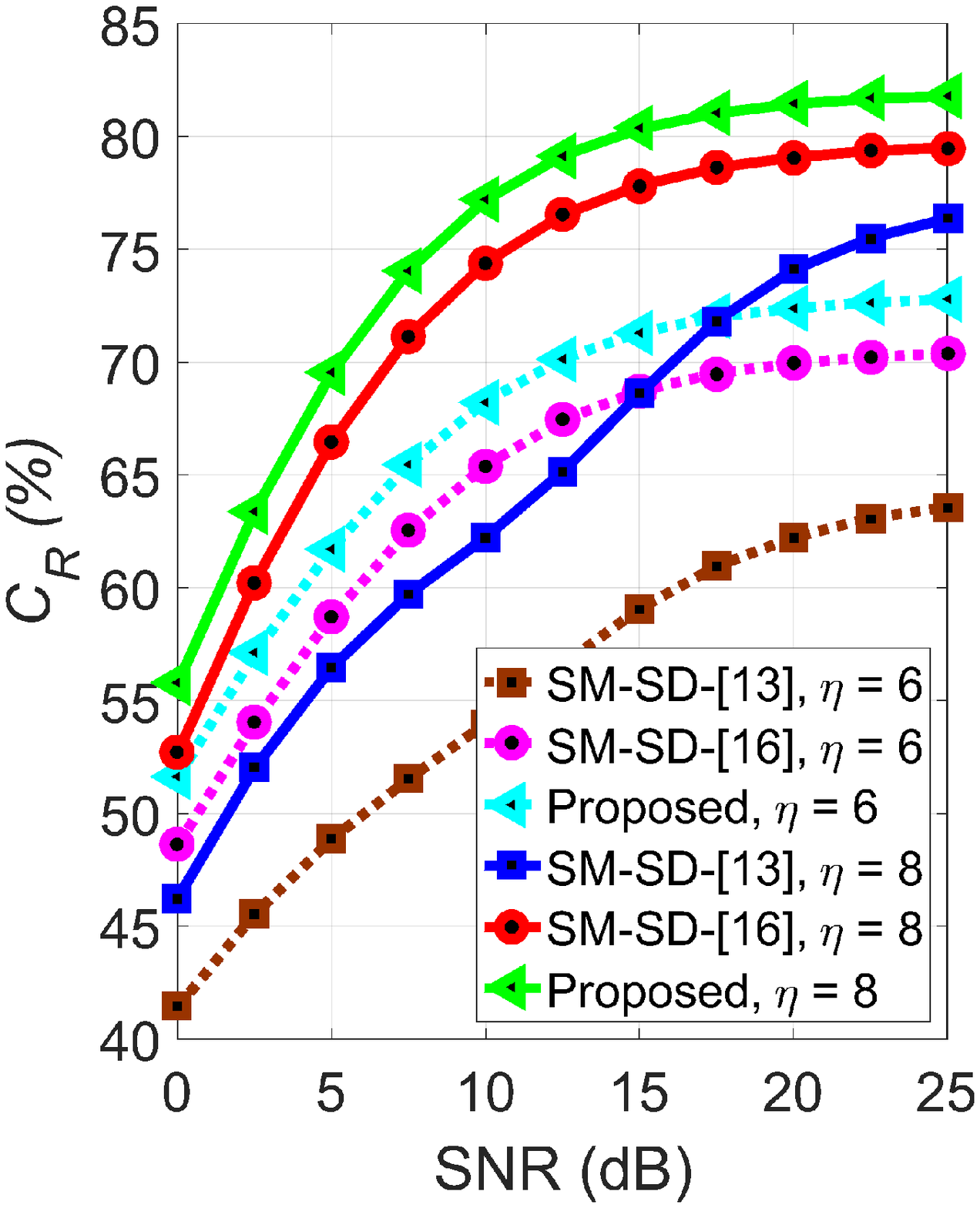}

}
\par\end{centering}
\caption{{\small{}\label{fig:Complexity-reduction-under}Complexity reduction
comparison of under-determined MIMO-SM system for different decoders.}}
\end{figure}

\begin{figure}
\begin{centering}
\subfloat[$\sigma_{e}^{2}=0$.]{\includegraphics[scale=0.31]{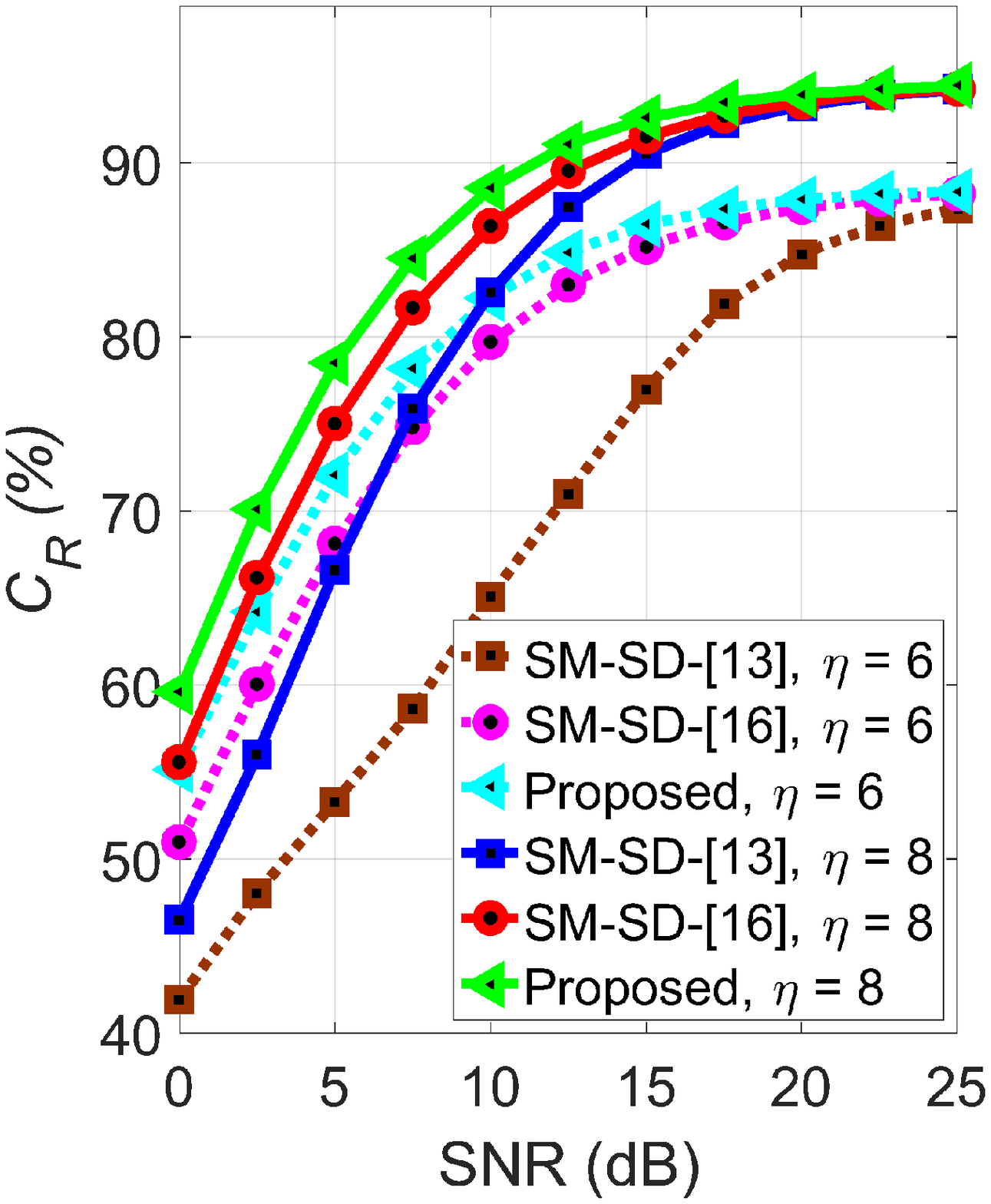}
\centering{}}~\subfloat[$\sigma_{e}^{2}=1/\text{snr}$.]{\includegraphics[scale=0.31]{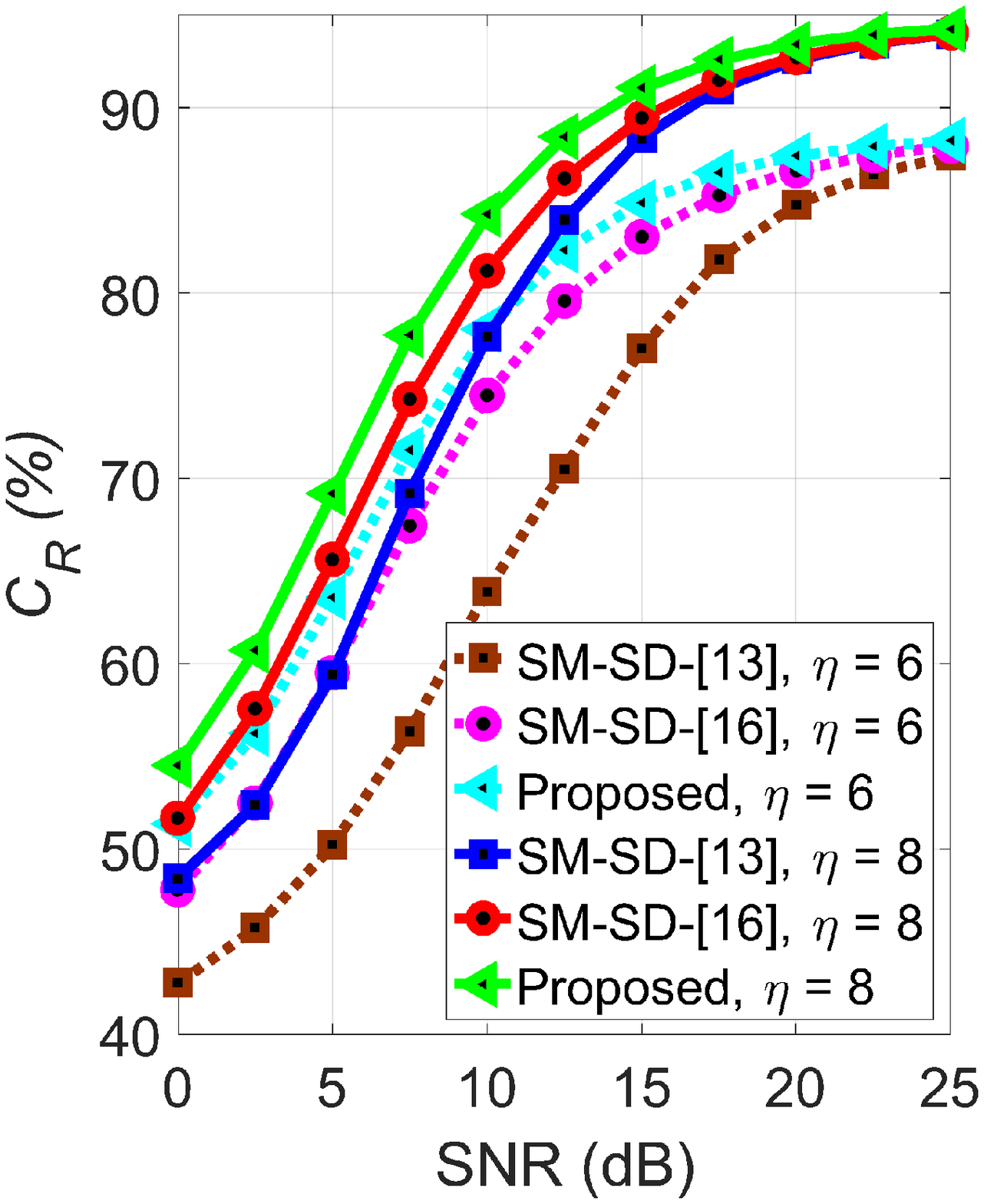}

}
\par\end{centering}
\begin{centering}
\subfloat[$\sigma_{e}^{2}=0.1$.]{\includegraphics[scale=0.31]{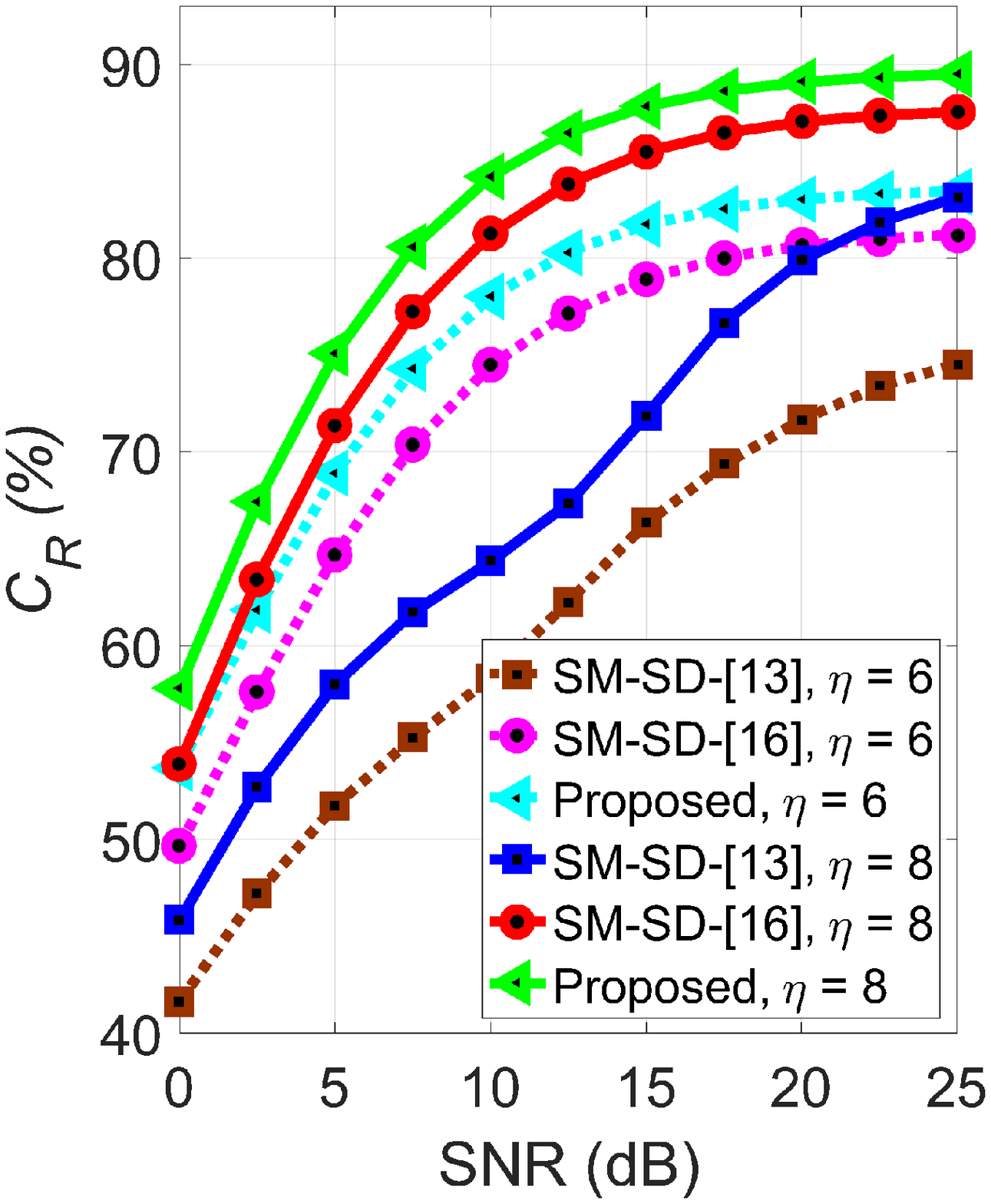}
\centering{}}~\subfloat[$\sigma_{e}^{2}=0.2$.]{\includegraphics[scale=0.31]{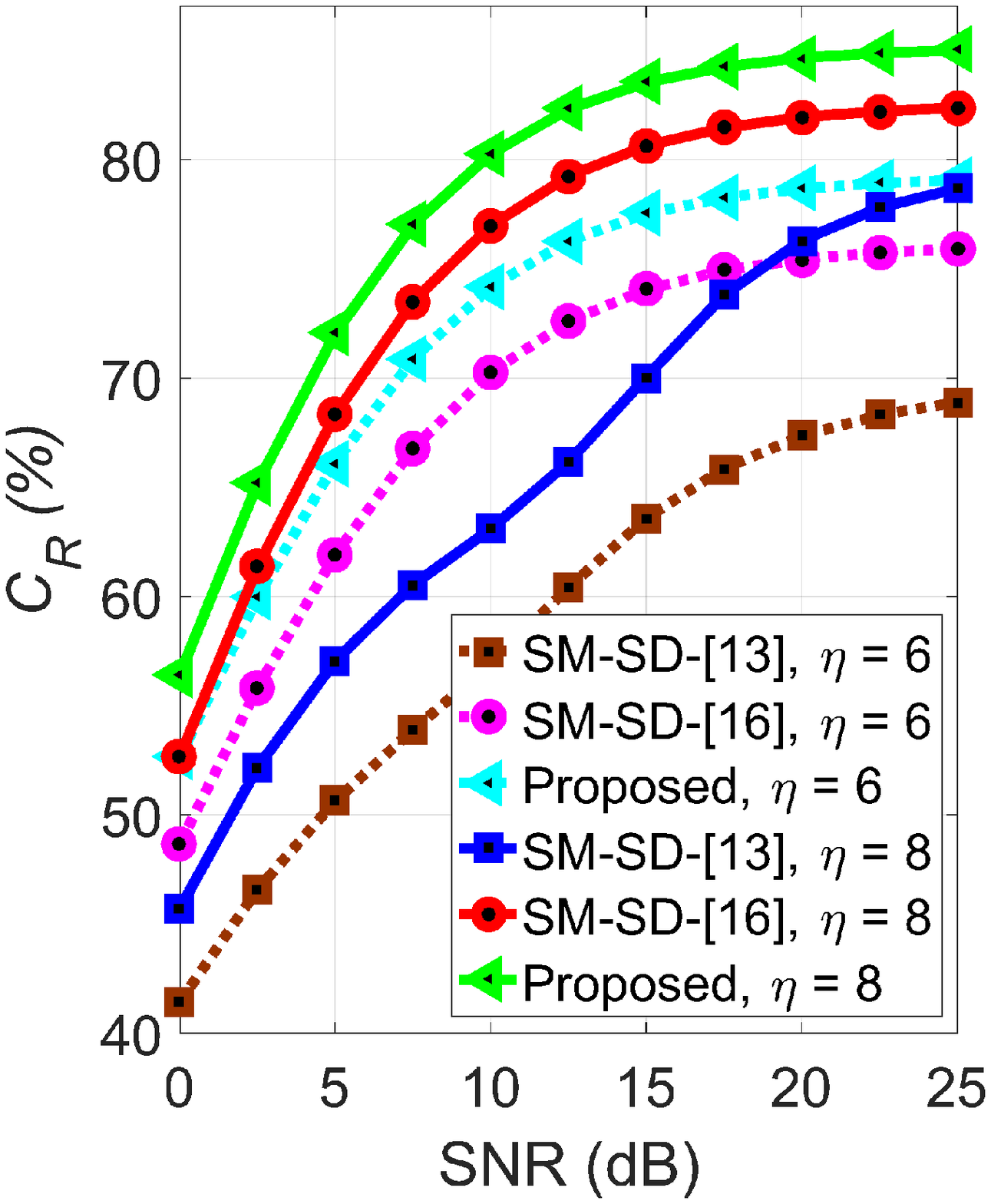}

}
\par\end{centering}
\caption{{\small{}\label{fig:Complexity-reduction-over}Complexity reduction
comparison of over-determined MIMO-SM system for different decoders.}}
\end{figure}

Figs. \ref{fig:Complexity-reduction-determined}, \ref{fig:Complexity-reduction-under}
and \ref{fig:Complexity-reduction-over} show the complexity reduction
ratio in (\ref{eq: Comp Reduction}) versus different values of SNR
for determined, under-determined, and over-determined MIMO-SM systems,
respectively. Each figure contains four sub-figures which represent
all scenarios of $\sigma_{e}^{2}$ (i.e., 0, 0.1, 0.2, and $1/$snr),
while each sub-figure presents the two available spectral efficiencies,
$\eta=6$ and $8$ bpcu. According to (\ref{eq: Comp Reduction max}),
$C_{\text{R}}^{\text{max}}=86.1\%$ and $93.4\%$ in Fig. \ref{fig:Complexity-reduction-determined}
for $\eta=6$ and $8$ bpcu, respectively; $C_{\text{R}}^{\text{max}}=82\%$
and $91.3\%$ in Fig. \ref{fig:Complexity-reduction-under} for $\eta=6$
and $8$ {\large{}bpcu,} respectively; and $C_{\text{R}}^{\text{max}}=88.6\%$
and $94.6\%$ in Fig. \ref{fig:Complexity-reduction-over} for $\eta=6$
and $8$ bpcu, respectively. 

In the case of $\sigma_{e}^{2}=0$ and $1/$snr, the proposed m-M
algorithm provides the best reduction in the decoding complexity ratio
over the SM-SD {[}\ref{A.-Younis,-S. IEEE Trans}{]} and SM-SD {[}\ref{I.-Al-Nahhal,-O. QSM}{]}
algorithms. The m-M algorithm as well as the other two algorithms
reach to $C_{\text{R}}^{\text{max}}$ at high SNR. It should be noted
that when $\eta$ increases, the decoding complexity ratio increases
for all algorithms. In the case of fixed $\sigma_{e}^{2}$ (i.e.,
0.1 and 0.2), no algorithm reaches $C_{\text{R}}^{\text{max}}$. However,
the proposed m-M algorithm provides the best reduction in the decoding
complexity ratio for all values of SNR. Also, as $\sigma_{e}^{2}$
increases, the reduction in complexity gain of the m-M algorithm over
the other two algorithm increases.

As it can be seen from these figures, the proposed m-M algorithm provides
a better complexity reduction ratio in the low SNR in the case of
perfect CSIR and variable $\sigma_{e}^{2}$. Moreover, it has the
superiority over the existing SM-SD algorithms for all values of SNR
in the case of imperfect CSIR with fixed $\sigma_{e}^{2}$. In addition,
the m-M algorithm is more robust to the increase of $\sigma_{e}^{2}$
than the existing SM-SD algorithms.

\subsection{Complexity Reduction Sensitivity}

\begin{figure}
\begin{centering}
\includegraphics[scale=0.35]{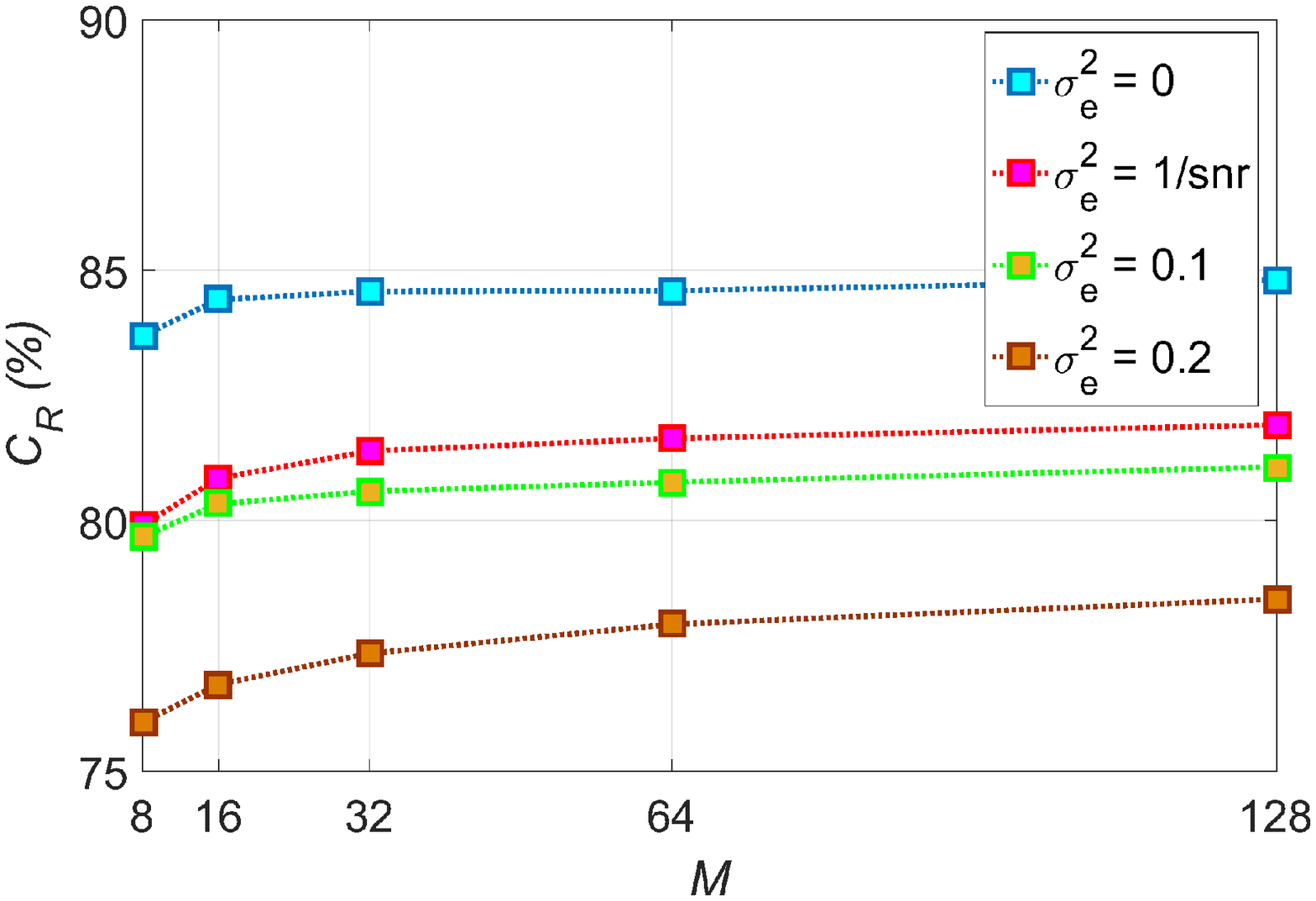}
\par\end{centering}
\caption{{\small{}\label{fig:Complexity-change M}Complexity reduction of the
proposed m-M algorithm for $N_{t}=N_{r}=16$ and variable $M$. }}
\end{figure}

\begin{figure}
\begin{centering}
\includegraphics[scale=0.35]{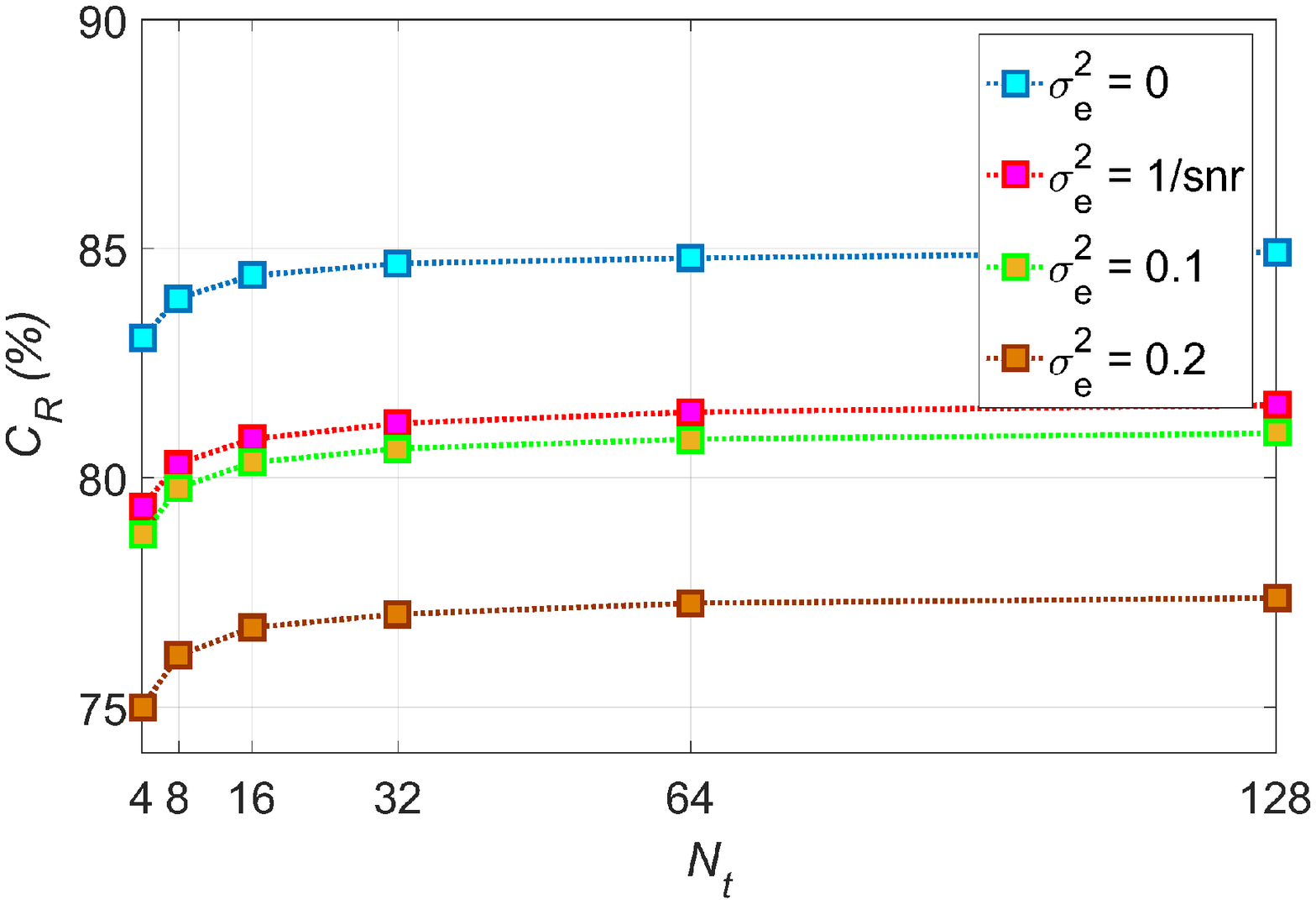}
\par\end{centering}
\caption{{\small{}\label{fig:Complexity-change Nt}Complexity reduction of
the proposed m-M algorithm for $M=N_{r}=16$ and variable $N_{t}$. }}
\end{figure}

\begin{figure}
\begin{centering}
\includegraphics[scale=0.35]{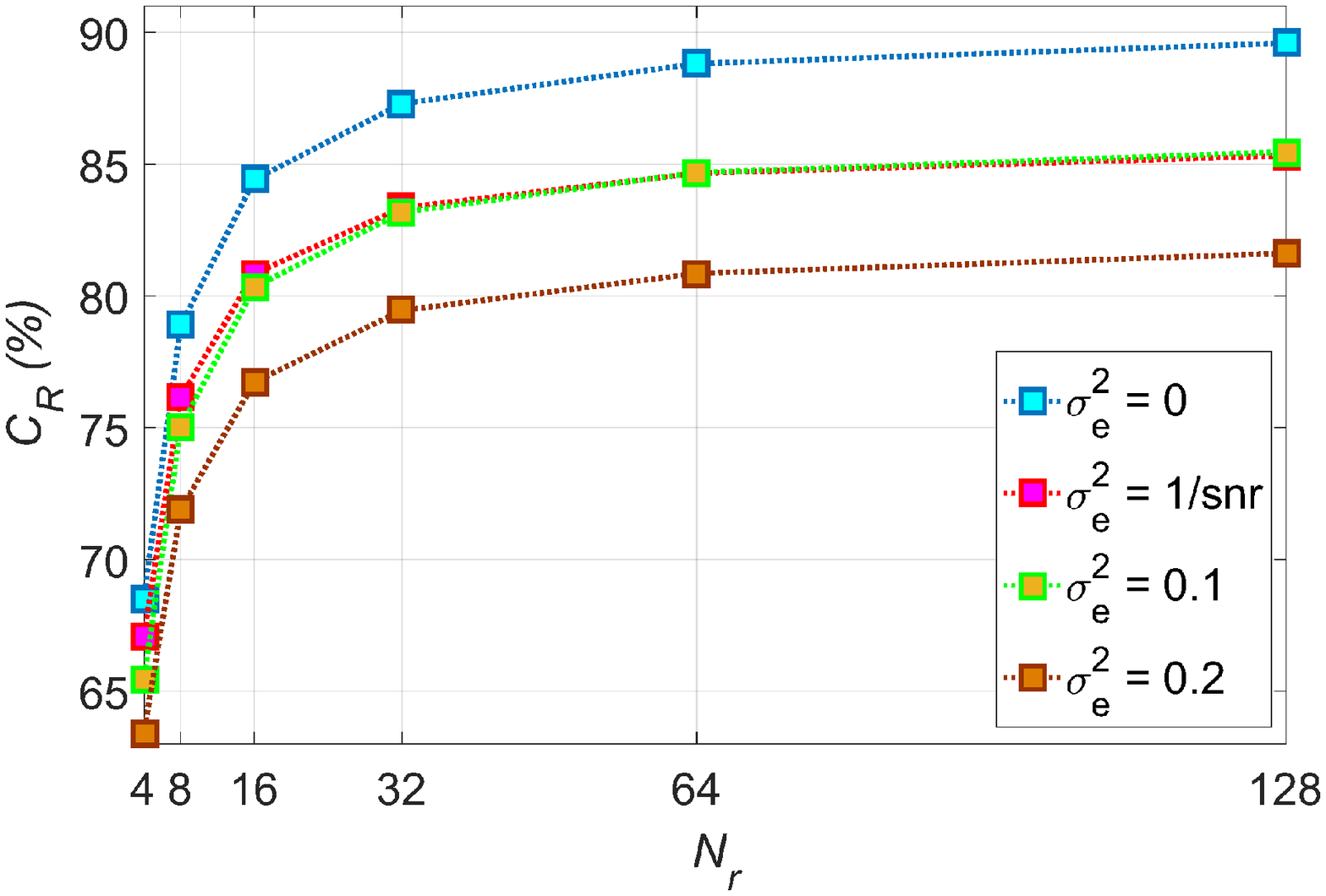}
\par\end{centering}
\caption{{\small{}\label{fig:Complexity-change Nr}Complexity reduction of
the proposed m-M algorithm for $N_{t}=M=16$ and variable $N_{r}$.}}
\end{figure}

We have noticed from Figs. \ref{fig:Complexity-reduction-determined},
\ref{fig:Complexity-reduction-under} and \ref{fig:Complexity-reduction-over}
that the reduction in the decoding complexity ratio for the m-M algorithm
increases as the MIMO-SM dimensions ($M$, $N_{t}$, and $N_{r}$)
increase. However, we need to determine which dimension affects more
the complexity reduction ratio. In this subsection, we assess the
reduction in the complexity ratio versus only one MIMO-SM dimension.

In Fig. \ref{fig:Complexity-change M}, the decoding complexity reduction
ratio of the m-M algorithm is assessed versus the QAM order, $M$,
for the $16\times16$ MIMO-SM system. It can be seen that the complexity
reduction ratio slightly increases as $M$ increases. For example,
for $\sigma_{e}^{2}=0.2$, the complexity reduction ratio increases
from $76\%$ to $78.5\%$ at $M=8$ and $128$, respectively. Thus,
the complexity reduction ratio of the proposed m-M algorithm is sensitive
to the slightly change of $M$.

In Fig. \ref{fig:Complexity-change Nt}, we evaluate the decoding
complexity reduction ratio of the m-M algorithm versus $N_{t}$ at
$N_{r}=M=16$. It can be noticed that the increase of the decoding
complexity reduction is negligible in comparison with the case of
variable $N_{t}$. Consequently, the change of $N_{t}$ has almost
no effect on the decoding complexity ratio of the m-M algorithm.

The decoding complexity reduction is evaluated versus different values
of $N_{r}$ in Fig. \ref{fig:Complexity-change Nr} for $N_{t}=M=16$.
We can see from this figure that the complexity reduction ratio increases
from $68\%$ at $N_{r}=4$ to $90\%$ at $N_{r}=128$ for $\sigma_{e}^{2}=0$,
and from $64\%$ at $N_{r}=4$ to $82\%$ at $N_{r}=128$ for $\sigma_{e}^{2}=0.2$.
Thus, the decoding complexity of the m-M algorithm increases logarithmically
as $N_{r}$ increases.

Finally, we can see from these figures that the decoding complexity
reduction ratio of the m-M algorithm is sensitive to the change of
$N_{r}$, while is nonsensitive to the changes of $N_{t}$ or $M$.

\subsection{Discussions}

As seen from our comprehensive comparisons, the proposed m-M algorithm
provides significant reduction in the decoding complexity basically
without BER performance loss. For SM systems, compressive sensing
(CS)-based algorithms have been recently proposed in {[}\ref{CS_2014}{]}-{[}\ref{CS_2016}{]}
to provide sub-optimal BER performance with a reduction in the decoding
complexity. These CS-based algorithms exploit the sparsity of the
SM signals to provide low-complexity detection at the expense of BER
deterioration. Normally, the CS-based algorithms are suitable for
over-determined MIMO-SM systems (i.e., $N_{r}>N_{t}$) to reduce the
BER performance gap versus the ML solution. The authors of {[}\ref{CS_2016}{]}
have proposed an enhanced Bayesian CS (EBCS) algorithm to provide
low-complexity detection with near ML BER performance. The minimum
decoding complexity of the EBCS algorithm in {[}\ref{CS_2016}{]}
can be achieved at high SNR, which is about $\mathcal{O}(N_{r}N_{t}^{2})+\mathcal{O}(N_{r}N_{t})+\mathcal{O}(N_{t})+\mathcal{O}(N_{r})$
floating point operations (flops). Since the ML decoder costs $9MN_{r}N_{t}$
flops, the maximum complexity reduction that can be achieved from
{[}\ref{CS_2016}{]} when compared with the ML decoder in high SNR
is $87.2\%$ and $88.1\%$ for $12\times8$ MIMO-SM with 8-QAM and
$20\times16$ MIMO-SM with 16-QAM, respectively. As shown in Fig.
\ref{fig:Complexity-reduction-over}-(a) and (\ref{eq: Comp Reduction max}),
the proposed m-M algorithm provides $88.6\%$ and $94.6\%$ complexity
reduction after 15 dB for $12\times8$ MIMO-SM with 8-QAM and $20\times16$
MIMO-SM with 16-QAM, respectively. Thus, the proposed algorithm has
a higher complexity reduction without any BER performance loss when
compared with the ML decoder.

Another recent low-complexity algorithm that provides a near-ML BER
performance is proposed in {[}\ref{X.-Zhang,-Y.}{]} by dividing the
tree-search into $N_{t}$ subtrees with $2N_{r}$ levels (for the
real-form representation of (\ref{eq: x_est_ML_index})) and $M$
branches. The transmit and receive antennas are ordered to reach the
solution faster. In the first subtree, the algorithm visits a different
number of nodes in each level, $K=[k_{1}\,\,k_{2}\,\,\ldots k_{2N_{r}}]$,
where $k_{i}$ represents the number of best nodes that should be
kept in the $i$-th level and expanded in the next level. The minimum
ED at the final level is used as a pruned radius for scanning the
next $N_{t}-1$ subtrees by applying the SD concept in {[}\ref{A.-Younis,-R.GLOBOCOM}{]}.
In high SNR, the minimum decoding complexity of the algorithm in {[}\ref{X.-Zhang,-Y.}{]}
is $(\sum_{i=1}^{2N_{r}}k_{i})+M(N_{t}-1)$ visited nodes plus the
cost of Eq. (5) in {[}\ref{X.-Zhang,-Y.}{]}. As discussed in (\ref{eq: Comp Reduction max}),
the proposed m-M algorithm can visit only $(2N_{r}+MN_{t}-1)$ nodes
to achieve the optimum BER performance. For instance, for a $4\times4$
MIMO-SM system with 64-QAM and $K=[64\,\,26\,\,26\,\,8\,\,8\,\,2\,\,2\,\,1]$
as mentioned in {[}\ref{X.-Zhang,-Y.}{]}, the minimum decoding complexity
of {[}\ref{X.-Zhang,-Y.}{]} in high SNR is $329$ visited nodes plus
the cost of Eq. (5) in {[}\ref{X.-Zhang,-Y.}{]}, while our proposed
algorithm visits only 263 nodes to achieve the optimum BER performance
in high SNR (almost high SNR is after 15 dB, as shown in Figs. \ref{fig:Complexity-reduction-determined},
\ref{fig:Complexity-reduction-under} and \ref{fig:Complexity-reduction-over}).
Thus, the m-M algorithm provides a lower decoding complexity than
the algorithm in {[}\ref{X.-Zhang,-Y.}{]} without losing the optimality
of BER performance.

For high rate SM transmissions, one of the suggested solutions is
to use a high value of $N_{t}$. Two systems are proposed to provide
high rate transmission using smaller $N_{t}$; 1) generalized SM (GSM)
which activates more than one transmit antenna at a time {[}\ref{GSM}{]},
and 2) quadrature SM (QSM) which delivers the symbols using the in-phase
and quadrature dimensions {[}\ref{QSM}{]}. At the receiver side,
the GSM and QSM systems have a similar tree-search structure to the
SM, and hence, the proposed m-M algorithm can be applied in a straightforward
manner.

\section{\label{sec:Conclusion}Conclusion}

This paper has proposed a novel low-complexity decoding algorithm
for MIMO-SM systems, referred to as the m-M algorithm. The m-M algorithm
provides a significant reduction in the decoding complexity in terms
of the number of nodes which are visited during the algorithm run.
The proposed algorithm guarantees achieving the ML solution by employing
a single expansion to the minimum ED across all  tree-search branches,
and stopping if this minimum ED occurs at the end of a fully expanded
tree-search branch. Furthermore, tight expressions for the expected
decoding complexity of the m-M algorithm have been derived. The proposed
algorithm and analytical expressions have been assessed in three different
scenarios: perfect CSIR, as well as imperfect CSIR with a fixed and
a variable channel estimation error variances, respectively. All scenarios
have been investigated for different types of MIMO-SM systems including
determined, under-determined, and over-determined systems. The numerical
results have shown that the proposed algorithm provides the best reduction
in the decoding complexity over existing optimal SM-SD algorithms.
The future work may focus on the development of the soft-decoding
version of the m-M algorithm.

\end{document}